\documentclass[10pt,english,
]{csthesis}

\usepackage{s5paper}
\usepackage[english]{babel}
\usepackage{titling}

\usepackage{amsmath}
\usepackage{amsfonts}
\usepackage{booktabs}

\renewcommand*\thesection{\arabic{section}}

\usepackage[vlined, ruled, lines numbered]{algorithm2e}

\providecommand{\sabstract}[1]{\textbf{\textit{Abstract }} #1}
\providecommand{\keywords}[1]{\textbf{\textit{Keywords }} #1}

\usepackage{tikz}
\usetikzlibrary{shapes,snakes,positioning}
\usetikzlibrary{arrows}
\usetikzlibrary{decorations.markings}
\usetikzlibrary{patterns}

\usepackage{tikz}
\usetikzlibrary{decorations.pathreplacing,calc}

\pretitle{\begin{center}\Huge\bfseries}
\posttitle{\par\end{center}\vskip 0.5em}
\preauthor{\begin{center}\Large\ttfamily}
\postauthor{\end{center}}
\predate{\par\large\centering}
\postdate{\par}

\begin{document}



\begin{center}
\LARGE{Robust level-3 BLAS Inverse Iteration from the Hessenberg Matrix}\\[3ex]
\end{center}

\begin{center}
\large{Angelika Schwarz (angies@cs.umu.se)\\[1ex]
Department of Computing Science, Ume\r{a} University, Sweden\\[1ex]
\today}
\end{center}

\sabstract{Inverse iteration is known to be an effective method for computing eigenvectors corresponding to simple and well-separated eigenvalues. In the non-symmetric case, the solution of shifted Hessenberg systems is a central step. Existing inverse iteration solvers approach the solution of the shifted Hessenberg systems with either RQ or LU factorizations and, once factored, solve the corresponding systems. This approach has limited level-3 BLAS potential since distinct shifts have distinct factorizations. This paper rearranges the RQ approach such that data shared between distinct shifts is exposed. Thereby the backward substitution with the triangular R factor can be expressed mostly with matrix--matrix multiplications (level-3 BLAS). The resulting algorithm computes eigenvectors in a tiled, overflow-free, and task-parallel fashion. The numerical experiments show that the new algorithm outperforms existing inverse iteration solvers for the computation of both real and complex eigenvectors.}\\

\keywords{inverse iteration, shifted Hessenberg systems, overflow-free computation}


\section{Introduction}\label{sec:introduction}
Inverse iteration is an established method for computing eigenvectors. When an approximation $\lambda$ to an eigenvalue of a matrix $\boldsymbol{A}$ is known, inverse iteration approximates an eigenvector by solving
\begin{displaymath}
(\boldsymbol{A} - \lambda \boldsymbol{I}) \boldsymbol{z}^{(k)} = \rho^{(k)} \boldsymbol{z}^{(k-1)} \qquad k \geq 1.
\end{displaymath}
Here, $\boldsymbol{z}^{(0)}$ is a unit norm starting vector and $\rho^{(k)}$ is a scalar that normalizes the iterate $\boldsymbol{z}^{(k)}$.
A converging sequence of $\boldsymbol{z}^{(k)}$ yields a right eigenvector $\boldsymbol{z} \neq \boldsymbol{0}$ that is an exact eigenvector of a nearby matrix $\boldsymbol{A}+\boldsymbol{E}$ with $||\boldsymbol{E}|| = \mathcal{O}(\epsilon ||\boldsymbol{A}||)$, where $\epsilon$ denotes the machine precision, implying a small residual $||\boldsymbol{A} \boldsymbol{z} - \lambda \boldsymbol{z}|| = \mathcal{O}(\epsilon ||\boldsymbol{A}||)$.

This papers concerns the case when $\boldsymbol{A}$ is non-symmetric and real. A standard approach reduces $\boldsymbol{A}$ to an upper Hessenberg matrix $\boldsymbol{H} = \boldsymbol{Q}_0^T \boldsymbol{A} \boldsymbol{Q}_0$, where $\boldsymbol{Q}_0$ is an orthogonal matrix. Then inverse iteration approximates an eigenvector $\boldsymbol{x} \neq \boldsymbol{0}$ of $\boldsymbol{H}$ by
\begin{equation}\label{eq:inviteration}
(\boldsymbol{H} - \lambda \boldsymbol{I}) \boldsymbol{x}^{(k)} = s^{(k)} \boldsymbol{x}^{(k-1)} \qquad k \geq 1.
\end{equation}
The scalar $s^{(k)}$ is again chosen so that $\boldsymbol{x}^{(k)}$ has unit norm. A computed eigenvector $\boldsymbol{x}$ is backtransformed into an eigenvector of $\boldsymbol{A}$ by $\boldsymbol{z} \gets \boldsymbol{Q}_0 \boldsymbol{x}$. Varah~\cite{varah1968calculation} showed that the starting vector $\boldsymbol{x}^{(0)}$ can be chosen such that a single iteration of \eqref{eq:inviteration} suffices. Then the task of computing an eigenvector $\boldsymbol{z}$ necessitates the efficient solution of shifted Hessenberg systems. This will be the topic of this paper.

Inverse iteration \eqref{eq:inviteration} hinges on the availability of good approximations $\lambda$ to the true eigenvalues. These approximations can be computed through the QR algorithm \emph{without} accumulating the orthogonal transformations~\cite[Section 7.6.1]{golub1996matrix}. This approach is indeed supported by the LAPACK 3.9.0 inverse iteration routine \textsc{DHSEIN}~\cite{anderson1999lapack}. 
If the Hessenberg matrix has zero subdiagonal elements, the eigenproblem decouples into smaller block-triangular problems. When the QR algorithm processes the blocks separately, \textsc{DHSEIN} exploits the known affiliation of the eigenvalue and block it belongs to and performs inverse iteration on the relevant block.
This paper is therefore based on two assumptions. First, good approximations to the eigenvalues of $\boldsymbol{H}$ are available. In other words, each approximation $\lambda$ is an exact eigenvalue of some matrix $\boldsymbol{H} + \boldsymbol{F}$, where $||\boldsymbol{F}|| = \mathcal{O}(\epsilon)$.
Second, the Hessenberg matrix $\boldsymbol{H}$ is unreduced. That is, all subdiagonal entries of $\boldsymbol{H}$ are non-zeros.

This paper offers a new algorithm for a more efficient computation of \eqref{eq:inviteration} if a batch of eigenvectors corresponding to distinct eigenvalues is sought. This new algorithm combines two ideas. The first idea is due to Henry~\cite{henry1994techreport} who addresses the solution of a shifted Hessenberg system through an RQ factorization. In a single sweep, $\boldsymbol{H}$ is reduced column-by-column to an upper triangular $\boldsymbol{R}$ such that the newly computed column of $\boldsymbol{R}$ is immediately used in the backward substitution. The level-3 BLAS potential in this approach is limited since distinct shifts results in distinct RQ factorizations. This problem has been addressed by Bosner et al.~\cite[Section~3]{bosner2018parallel} and gives the second idea. Level-3 BLAS can be introduced in spite of distinct shifts. A \emph{partially} computed (tiled) RQ factorization exposes data shared between distinct shifts. The computation can be arranged such that most of the computation correspond to matrix--matrix multiplications. Specifically, this paper contributes an inverse iteration algorithm based on solving shifted Hessenberg systems through the RQ approach with the following highlights.
\begin{itemize}
\item The RQ approach is revised such that most of the backward substitution corresponds to matrix--matrix multiplications (level-3 BLAS~\cite{BLAS}) in spite of distinct eigenvalues.
\item The new algorithm is tiled and can naturally be parallelized with tasks.
\item In existing inverse iteration solvers, complex eigenvector are significantly more expensive to compute than real eigenvectors. The new algorithm solves this issue and supports the computation of real and complex eigenvectors alike, meaning that the computational cost per column is approximately the same.
\end{itemize}

The rest of this paper is organized as follows. Section~\ref{sec:related-work} reviews the LU approach realized in \textsc{DHSEIN} and the RQ approach by Henry~\cite{henry1994techreport} to solving shifted Hessenberg systems as well as the tiled RQ factorization by Bosner et al.~\cite{bosner2018parallel}. Inspired by these ideas, Section~\ref{sec:main} presents the new algorithm for solving shifting Hessenberg system, which is the core of the inverse iteration routine developed in Section~\ref{sec:inverse-iteration}. Section~\ref{sec:experiments} describes the numerical experiments and presents their results.

\section{Related Work}\label{sec:related-work}
This section reviews approaches to the solution of shifted Hessenberg systems. First, the LU approach implemented in LAPACK and the RQ factorization advocated by Henry~\cite{henry1994techreport} are discussed. Then, the tiled RQ factorization by Bosner et al.~\cite{bosner2018parallel} delivers key ideas used in the new algorithm presented in Section~\ref{sec:main}.

\subsection{LU Factorization}\label{sec:LU}
LAPACK contains the inverse iteration routine \textsc{DLAEIN} \cite{anderson1999lapack} for computing a single eigenvector. \textsc{DLAEIN} addresses the solution of the first iteration of \eqref{eq:inviteration} through an LU factorization with partial pivoting
\begin{displaymath}
\boldsymbol{P}(\boldsymbol{H} - \lambda \boldsymbol{I}) = \boldsymbol{L}\boldsymbol{U}, \quad \boldsymbol{L} \boldsymbol{y} = \boldsymbol{P}\boldsymbol{x}^{(0)}, \quad \boldsymbol{U} \boldsymbol{x}^{(1)} = \alpha \boldsymbol{y}, \quad \boldsymbol{x} \gets s^{(1)} \boldsymbol{x}^{(1)}.
\end{displaymath}
Here, $\boldsymbol{U}$ is upper triangular, $\boldsymbol{L}$ is lower unit triangular and  $\boldsymbol{P}$ is a permutation matrix. As the starting vector can be chosen such that a single iteration suffices~\cite{varah1968calculation,peters1971calculation}, \textsc{DLAEIN} executes only one half iteration solving $\boldsymbol{U} \boldsymbol{x}^{(1)} = \alpha \boldsymbol{y}$. The scalar $\alpha$ serves the avoidance of overflow. The vector $\boldsymbol{y}$ is set to a scaled vector of ones $\boldsymbol{e}$, where the scaling depends on $\boldsymbol{H}$. In other words, the starting vector is implicitly selected as $\boldsymbol{x}^{(0)} = \boldsymbol{P}^{-1} \boldsymbol{L} \boldsymbol{e}$. If the first initial vector $\boldsymbol{y}$ does not satisfy the convergence criterion in the first half iteration, the initial vector is changed rather than computing more iterations. Distinct shifts yield distinct LU factorizations of $\boldsymbol{H}-\lambda \boldsymbol{I}$. Hence, when several eigenvectors corresponding to distinct eigenvalues are sought, a different LU factorization is computed for each eigenvalue. To not overwrite $\boldsymbol{H}$, the upper triangular factor $\boldsymbol{U}$ is computed in a workspace.

\subsection{RQ and UL Factorization}\label{sec:rq}

\begin{algorithm}[t]
  \caption{Solving $(\boldsymbol{H} - \lambda \boldsymbol{I}) \boldsymbol{x} = \boldsymbol{b}$ with an RQ decomposition (Henry~\cite[Algorithm 2]{henry1994techreport})}
  \label{alg:henry}
  $\boldsymbol{v} \gets \boldsymbol{h}(1:n,n)$; $v(n) \gets v(n) - \lambda$; $\boldsymbol{x} \gets \boldsymbol{b}$\;
  \For{$k \gets n:-1:2$}{
  Determine a Givens rotation $\boldsymbol{G}^T_k = \begin{bmatrix}
  c_k & -s_k \\
  s_k & c_k
  \end{bmatrix}$ such that $\begin{bmatrix}  h(k,k-1) & v(k) \end{bmatrix} \boldsymbol{G}^T_k = \begin{bmatrix} 0 & \phi  \end{bmatrix}$\;  
  $x(k) = x(k) / \phi$\;  
  $\tau_1 \gets s_k x(k)$; $\tau_2 \gets c_k x(k)$\;
  $\boldsymbol{x}(1:k-2) \gets \boldsymbol{x}(1:k-2) - \tau_2 \boldsymbol{v}(1:k-2) + \tau_1 \boldsymbol{h}(1:k-2,k-1)$\;
  $x(k-1) \gets x(k-1) - \tau_2 v(k-1) + \tau_1 (h(k-1,k-1)-\lambda)$\;
  $\boldsymbol{v}(1:k-2) \gets c_k \boldsymbol{h}(1:k-2,k-1) + s_k \boldsymbol{v}(1:k-2)$\;
  $v(k-1) \gets c_k (h(k-1,k-1) - \lambda) + s_k v(k-1)$\;
  }
  $\tau_1 \gets x(1)/v(1)$\;
  
  \For{$k \gets 2:n$}{
  		$\tau_2 \gets x(k)$\;
  		$x(k-1) \gets c_k \tau_1 - s_k \tau_2$\;
  		$\tau_1 \gets c_k \tau_2 + s_k \tau_1$\;
  }
  \Return $\boldsymbol{y}$\;
  
\end{algorithm}

Henry~\cite{henry1994techreport,henry1995parallel} approaches the solution of $(\boldsymbol{H} - \lambda \boldsymbol{I}) \boldsymbol{x} = \boldsymbol{b}$ through an RQ factorization. By applying suitable Givens rotations from the right,
\begin{align*}
(\boldsymbol{H} - \lambda \boldsymbol{I}) \underbrace{\boldsymbol{G}_n^T \boldsymbol{G}_{n-1}^T \hdots \boldsymbol{G}_2^T}_{\boldsymbol{Q}^T} \underbrace{\boldsymbol{G}_2 \hdots \boldsymbol{G}_{n}}_{\boldsymbol{Q}} \boldsymbol{x} = \boldsymbol{b},
\end{align*}
the shifted Hessenberg matrix is transformed into an upper triangular matrix $\boldsymbol{R} = (\boldsymbol{H} - \lambda \boldsymbol{I}) \boldsymbol{Q}^T$. The solution $\boldsymbol{x}$ is obtained by solving $\boldsymbol{R y} = \boldsymbol{b}$ and backtransforming $\boldsymbol{x} = \boldsymbol{Q}^T \boldsymbol{y}$. By applying the Givens rotations, the columns of $\boldsymbol{R}$ are successively computed from right to left. For this purpose, the Givens rotation $\boldsymbol{G}_k^T$ is constructed to annihilate the subdiagonal entry $h(k,k-1)$ and transforms the columns $k-1$ and $k$. As soon as a column of $\boldsymbol{R}$ has been computed, it is immediately used in a column-wise backward substitution and then discarded. Henry uses an auxiliary column to compute and store this single column of $\boldsymbol{R}$. The full procedure is listed in Algorithm~\ref{alg:henry}. The flop count of Algorithm~\ref{alg:henry} is $3.5n^2 + \mathcal{O}(n)$ for a real shift. By virtue of merging the computation of the R factor and the backward substitution, the matrix $\boldsymbol{H}$ is accessed only once. Since $\boldsymbol{H}$ is left untouched, several shifted Hessenberg systems can be solved simultaneously and benefit from improved temporal locality of accessing columns of $\boldsymbol{H}$.

When the shift in Algorithm~\ref{alg:henry} is complex, both the R and the Q factor become complex. The backward substitution with R then relies fully on complex arithmetic and multiplies complex vectors with complex scalars. To exploit that all entries but the diagonal of $\boldsymbol{H}-\lambda \boldsymbol{I}$ are real, Henry~\cite[Section 4]{henry1994techreport} employs the RQ approach only for real shifts. Complex shifts, by contrast, are addressed by an UL factorization. The UL approach can benefit from mixed real-complex arithmetic such that real vectors are multiplied with complex scalars. Depending on what Gauss transformation is used, the UL approach requires $3n^2 + \mathcal{O}(n)$ or $3.5n^2 + \mathcal{O}(n)$ flops.

\subsection{Tiled RQ Factorization}\label{sec:tiled-rq}

Bosner, Bujanovi\'{c} and Drma{\v{c}} \cite{bosner2018parallel} generalize Henry's RQ factorization. They target the computation of the complex frequency response function (transfer function) $\boldsymbol{G}(\sigma_\ell) = \boldsymbol{C} ( \boldsymbol{A} - \sigma_\ell \boldsymbol{I})^{-1} \boldsymbol{B}$ through a reduction to the controller-Hessenberg form $\tilde{\boldsymbol{C}} (\tilde{\boldsymbol{A}} - \sigma_\ell \boldsymbol{I})^{-1} \tilde{\boldsymbol{B}}$. Here, $\tilde{\boldsymbol{A}} = \boldsymbol{Q}^H \boldsymbol{A} \boldsymbol{Q} \in \mathbb{C}^{n \times n}$ is m-Hessenberg (Hessenberg with m subdiagonals), $\tilde{\boldsymbol{C}} = \boldsymbol{C} \boldsymbol{Q} \in \mathbb{C}^{n \times m}$, $\tilde{\boldsymbol{B}} = \boldsymbol{Q}^H \boldsymbol{B} \in \mathbb{C}^{p \times n}$ is upper triangular in the first $p$ rows and otherwise filled with zeros, and $\boldsymbol{Q} \in \mathbb{C}^{n \times n}$ is unitary. A solution can be obtained by substituting the RQ factorization $\tilde{\boldsymbol{A}} - \sigma_\ell \boldsymbol{I} = \boldsymbol{R}_\ell \boldsymbol{Q}_\ell$ into the controller-Hessenberg form. Then $\tilde{\boldsymbol{C}} (\boldsymbol{R}_\ell \boldsymbol{Q}_\ell)^{-1} \tilde{\boldsymbol{B}} = (\tilde{\boldsymbol{C}} \boldsymbol{Q}^H) (\boldsymbol{R}_{\ell}^{-1} \tilde{\boldsymbol{B}})$ can be evaluated by solving a triangular system and two matrix--matrix multiplications.

The authors devise a tile column-oriented algorithm that computes RQ factorizations simultaneously for distinct shifts. For each $\ell = 1,\hdots,s$, the currently processed tile column with $m+n_{\rm{b}}$ columns is split into a window on the diagonal and the remaining offdiagonal part. The window is $n_{\rm{b}}$-by-$(m + n_{\rm{b}})$ and m-Hessenberg. It is reduced to triangular shape with $m$ leading zero columns by applying unitary transformations from the right. The unitary transformations are \emph{only} applied to the window and are accumulated into a matrix $\boldsymbol{Q}_{\ell}$. It remains to update the offdiagonal part with $\boldsymbol{Q}_{\ell}$. The offdiagonal part is partitioned $\begin{bmatrix}
\boldsymbol{D} & \boldsymbol{E}_{\ell}
\end{bmatrix}$. The matrix $\boldsymbol{D}$ is shared across shifts because the handling of the differences due to the shift-specific diagonal entries is deferred to a correction step. The matrix $\boldsymbol{E}_{\ell}$ holds the shift-distinct columns that have been transformed with previous unitary transformations. The update $\begin{bmatrix}
\boldsymbol{D} & \boldsymbol{E}_{\ell}
\end{bmatrix} \boldsymbol{Q}_{\ell}^H(1:n_{\rm{b}}+m,:)$ is split into two parts. First, the so-called batched GEMM processes the shift-specific matrix--matrix multiplications $\boldsymbol{E}_{\ell} \boldsymbol{Q}_{\ell}(n_{\rm{b}}+1:n_{\rm{b}}+m,:)$. Second, the shared $\boldsymbol{D}$ is exploited and updated with a joint matrix--matrix multiplication 
$\boldsymbol{D} \begin{bmatrix}
\boldsymbol{Q}_1^H(1:n_{\rm{b}},:) & \boldsymbol{Q}_2^H(1:n_{\rm{b}},:) &\hdots & \boldsymbol{Q}_s^H(1:n_{\rm{b}},:)
\end{bmatrix}$.

The tiled RQ approach delivers two key concepts that are used for the algorithm developed in the next section. First, the algorithm is designed as a tile column-oriented algorithm such that the shift-specific orthogonal transformations are only applied to the small window on the diagonal. Second, working with a partially reduced Hessenberg matrix and rearranging the computation exposes data that is shared between distinct shift. These two concepts combined Henry's column-oriented backward substitution algorithm introduce level-3 BLAS potential.

\section{Solution of Shifted Hessenberg Systems with Level-3 BLAS}\label{sec:main}
In this section we devise a tile column-oriented algorithm that solves $(\boldsymbol{H} - \lambda_\ell \boldsymbol{I})\boldsymbol{x}_\ell = \boldsymbol{b}_{\ell}$ simultaneously for many distinct shifts $\lambda_{\ell}$. Using ideas by Bosner et al.~\cite{bosner2018parallel}, we adapt the RQ approach by Henry~\cite{henry1994techreport} such that a large part of the data is shared. This way, the backward substitution phases can be merged for several shifts such that a large fraction of the computation corresponds to matrix--matrix multiplications.

\subsection{Simultaneous Backward Substitution of a Batch of Shifts}
The RQ approach requires a different sequence of Givens rotations for every shift $\lambda_{\ell}$, $\ell = 1, \hdots, m$. For simplicity, we assume $\lambda_{\ell}$ to be real and defer the complex case to Section~\ref{sec:complex-shifts}. Let
\begin{equation}\label{eq:givens}
\boldsymbol{G}_{j,\ell}^T =
\left[\begin{array}{cccc}
\boldsymbol{I}_{j-2} & \boldsymbol{0} & \boldsymbol{0} & \boldsymbol{0} \\
\boldsymbol{0}       & c(j,{\ell})    & -s(j,{\ell})   & \boldsymbol{0} \\
\boldsymbol{0}       & s(j,{\ell})    &  c(j,{\ell})   & \boldsymbol{0} \\
\boldsymbol{0}       & \boldsymbol{0} & \boldsymbol{0} & \boldsymbol{I}_{n-j}
\end{array}\right] \in \mathbb{R}^{n \times n}
\end{equation}
be the Givens rotation that transforms the columns $j$ and $j-1$ of $\boldsymbol{H} - \lambda_{\ell}\boldsymbol{I}$. To be available in the backtransformation, the cosine and sine components of all Givens rotations are recorded in the matrices $\boldsymbol{C} = [c(j,\ell)]\in \mathbb{R}^{n \times m}$ and $\boldsymbol{S}=[s(j,\ell)]\in \mathbb{R}^{n \times m}$.

The RQ approach transforms the Hessenberg matrix $(\boldsymbol{H} - \lambda_{\ell} \boldsymbol{I}) \boldsymbol{G}_{n,\ell}^T \hdots \boldsymbol{G}_{2,\ell}^T = \boldsymbol{R}_{\ell}$. We apply the Givens rotations in batches and thereby compute a triangular factor $\boldsymbol{R}_{\ell}$ tile column by tile column starting from the right.

\begin{figure}[h!]
\centering
\begin{tikzpicture}[scale=1.25]

\draw[](0,0) -- (3,0) -- (3,-3) -- cycle;

\draw[](0,-0.1) -- (2.9,-3);


\draw[](1,0) -- (1,-1.1);
\draw[](2,0) -- (2,-2.1);

\draw[](1.9,-2) -- (3,-2);
\draw[](0.9,-1) -- (2,-1);

\draw [decorate,decoration={brace,amplitude=10pt,mirror,raise=4pt},yshift=0pt](3,0.00) -- (2,0.0) node[black,midway,xshift=0.cm,yshift=0.7cm] {$k:n$};

\draw [decorate,decoration={brace,amplitude=10pt,mirror,raise=4pt},yshift=0pt](8,0.0) -- (7,0.0) node[black,midway,xshift=0.cm,yshift=0.7cm] {$k:n$};

\node[]() at (4,-1.5){$\boldsymbol{G}_{n, \ell}^T \hdots \boldsymbol{G}_{k,\ell}^T = $};

\draw[](5,0) -- (8,0) -- (8,-3) -- cycle;

\draw[](5,-0.1) -- (6.9,-2);

\draw[fill=yellow](7,0) -- (7,-2.0) -- (6.9,-2.0) -- (6.9,0) -- cycle;
\node[](k) at (6.8,0.5){$\tilde{\boldsymbol{r}}_{\ell}$};\draw[->](6.8,0.4)--(6.95,-0.15);

\draw[fill=black!20](7,0) -- (7,-2.0) -- (8,-3.0) -- (8,0) -- cycle;

\draw[](6,0) -- (6,-1.1);
\draw[](7,0) -- (7,-2.);

\draw[](7,-2) -- (8,-2);
\draw[](5.9,-1) -- (7,-1);
\end{tikzpicture}
\caption{Reduction of the rightmost tile column.}\label{fig:reduction-rightmost-tile-column}
\end{figure}
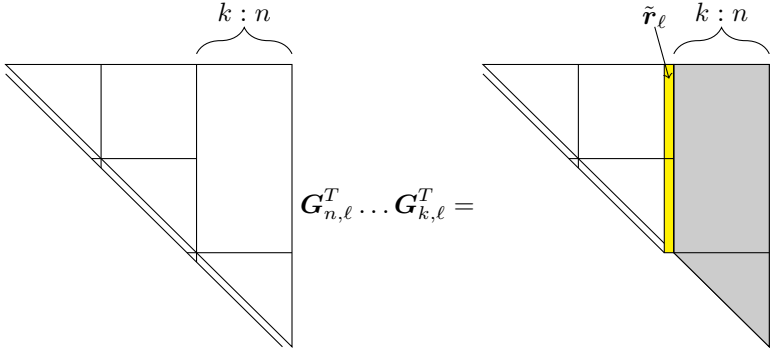

\emph{Rightmost tile column}: The first batch of Givens rotations $\boldsymbol{G}_{n,\ell}^T, \hdots, \boldsymbol{G}_{k,\ell}^T $ transforms the columns $k:n$ of $\boldsymbol{H}- \lambda_{\ell}\boldsymbol{I}$ into $\boldsymbol{R}_{\ell}(1:n,k:n)$, see Figure~\ref{fig:reduction-rightmost-tile-column}. Matrix entries that are updated are highlighted in the figure. The Givens rotation $\boldsymbol{G}_{k,\ell}^T$ transforms the columns $k$ and $k-1$, where $k-1$ is outside of the currently processed tile column. Hence, applying $\boldsymbol{G}_{k,\ell}^T$ yields a cross-over column $\tilde{\boldsymbol{r}}_{\ell}$.

In particular, the diagonal tile has been transformed into a triangular matrix $\boldsymbol{R}_{\ell}(k:n,k:n)$. The triangular system $\boldsymbol{R}_{\ell}(k:n,k:n) \boldsymbol{x}_{\ell}(k:n) = \boldsymbol{b}_{\ell}(k:n)$ can be solved by backward substitution for every shift (level-2 BLAS). For brevity, let $\mathcal{K} := 1:k-1$ be the index range of the remaining row indices. In a tile column-oriented backward substitution algorithm, the computed solution is used to update $\boldsymbol{b}_{\ell}(\mathcal{K}) \gets \boldsymbol{b}_{\ell}(\mathcal{K}) - \boldsymbol{R}_{\ell}(\mathcal{K},k:n) \boldsymbol{x}_{\ell}(k:n)$. If $\boldsymbol{R}_{\ell}(\mathcal{K},k:n)$ is \emph{not} generated explicitly, the update reads
\begin{align*}
\boldsymbol{b}_{\ell}(\mathcal{K}) \gets \boldsymbol{b}_{\ell}(\mathcal{K}) -  \underbrace{\begin{bmatrix}
\boldsymbol{h}(\mathcal{K},k-1) - \lambda_{\ell} \boldsymbol{e}_{k-1} & \boldsymbol{H}(\mathcal{K},k:n)
\end{bmatrix} \boldsymbol{G}_1^T}_{ 
\begin{bmatrix}
\tilde{\boldsymbol{r}}_{\ell} & \boldsymbol{R}_{\ell}(\mathcal{K},k:n)
\end{bmatrix}
}
\begin{bmatrix}
0 \\
\boldsymbol{x}_{\ell}(k:n)
\end{bmatrix}
\end{align*}
where $\boldsymbol{G}^T_1 = \boldsymbol{G}_{n,\ell}^T(k-1:n,k-1:n) ...\boldsymbol{G}_{k,\ell}^T(k-1:n,k-1:n)$ and $\boldsymbol{e}_{k-1}$ is the vector whose $(k-1)$-th entry is one and all other entries are zero. The update can be bracketed so that the Givens rotations are applied to the right-hand side. Then the update
\begin{displaymath}
\boldsymbol{b}_{\ell}(\mathcal{K}) \gets \boldsymbol{b}_{\ell}(\mathcal{K}) - \begin{bmatrix}
\boldsymbol{h}(\mathcal{K},k-1) - \lambda_{\ell} \boldsymbol{e}_{k-1} & \boldsymbol{H}(\mathcal{K},k:n)
\end{bmatrix} \underbrace{\boldsymbol{G}_1^T 
\begin{bmatrix}
0 \\
\boldsymbol{x}_{\ell}(k:n)
\end{bmatrix}}_{\boldsymbol{z}_{\ell}}
\end{displaymath}
can be interpreted as a block matrix multiplication where the second block $\boldsymbol{H}(\mathcal{K},k:n)$ is shared across distinct shifts. Hence, when several right-hand sides $\boldsymbol{z}_{\ell}$ are computed simultaneously, the multiplication with the second block corresponds to a matrix--matrix multiplication (level-3 BLAS). Only the scalar-vector multiplication with the first block $\boldsymbol{h}(\mathcal{K},k-1)-\lambda_{\ell}\boldsymbol{e}_{k-1}$ is shift-specific. Note that the computation of $\boldsymbol{z}_{\ell}$ is cheap since every Givens rotation only transforms two entries of $\boldsymbol{x}_{\ell}$. 

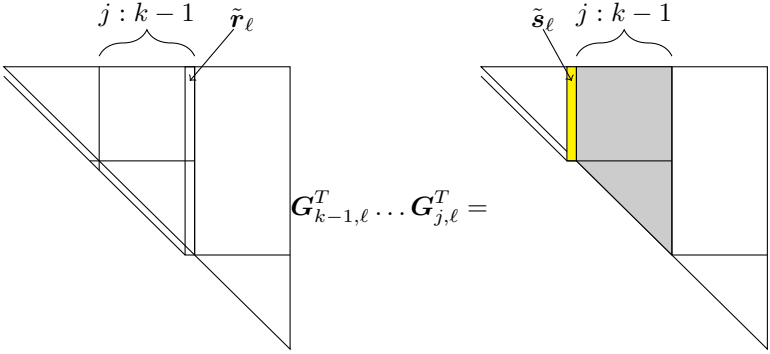
\begin{figure}[h!]
\centering
\begin{tikzpicture}[scale=1.25]

\draw[](0,0) -- (3,0) -- (3,-3) -- cycle;

\draw[](0,-0.1) -- (1.9,-2);

\draw[](2,0) -- (2,-2.0) -- (1.9,-2.0) -- (1.9,0) -- cycle;
\node[](k) at (2.5,0.5){$\tilde{\boldsymbol{r}}_{\ell}$};\draw[->](2.41,0.4)--(1.95,-0.15);

\draw[](1,0) -- (1,-1.1);
\draw[](2,0) -- (2,-2.);

\draw[](2,-2) -- (3,-2);
\draw[](0.9,-1) -- (2,-1);

\draw [decorate,decoration={brace,amplitude=10pt,mirror,raise=4pt},yshift=0pt](2,0.0) -- (1,0.0) node[black,midway,xshift=0.cm,yshift=0.7cm] {$j:k-1$};

\node[]() at (4,-1.5){$\;\boldsymbol{G}_{k-1,\ell}^T \hdots \boldsymbol{G}_{j,\ell}^T = $};

\draw[](5,0) -- (8,0) -- (8,-3) -- cycle;

\draw[](5,-0.1) -- (5.9,-1);

\draw [decorate,decoration={brace,amplitude=10pt,mirror,raise=4pt},yshift=0pt](7,0.0) -- (6,0.0) node[black,midway,xshift=0.cm,yshift=0.7cm] {$j:k-1$};

\draw[fill=yellow](6,0) -- (6,-1.0) -- (5.9,-1.0) -- (5.9,0) -- cycle;
\node[](k) at (5.65,0.5){$\tilde{\boldsymbol{s}}_{\ell}$};\draw[->](5.65,0.4)--(5.95,-0.15);

\draw[fill=black!20](6,0) -- (6,-1.0) -- (7,-2.0) -- (7,0) -- cycle;

\draw[](6,0) -- (6,-1.);
\draw[](7,0) -- (7,-2.);

\draw[](7,-2) -- (8,-2);
\draw[](5.9,-1) -- (7,-1);
\end{tikzpicture}
\caption{Reduction of a center tile column.}\label{fig:reduction-center-tile-column}
\end{figure}

\emph{Center tile columns:} In contrast to the rightmost tile column, center tile columns have shift-dependent cross-over columns and diagonal entries. The applications of the batch of Givens rotations $\boldsymbol{G}_{k-1,\ell}^T, \hdots,\boldsymbol{G}_{j,\ell}^T$ yields

\begin{align}\label{eq:center}
&\Bigg(\left[\begin{array}{c|cc}
\boldsymbol{h}(1:j-1,j-1) & \boldsymbol{H}(1:j-1,j:k-2) & \tilde{\boldsymbol{r}}_{\ell}(1:j-1) \\
\boldsymbol{h}(j:k-1,j-1) & \boldsymbol{H}(j:k-1,j:k-2) & \tilde{\boldsymbol{r}}_{\ell}(j:k-1) \\
\end{array}\right] \\
&\hspace{1cm}- \lambda_{\ell} 
\left[\begin{array}{c|c}
\boldsymbol{e}_{j-1} & \boldsymbol{0} \\
\boldsymbol{0}   & \boldsymbol{I} - \boldsymbol{e}_{k-1} \boldsymbol{e}_{k-1}^T
\end{array}\right]
\Bigg) \boldsymbol{G}_2^T =\left[\begin{array}{c|c}
\tilde{\boldsymbol{s}}_{\ell} & \boldsymbol{R}_{\ell}(1:j-1,j:k-1)  \\
\boldsymbol{0} & \boldsymbol{R}_{\ell}(j:k-1,j:k-1)  \\
\end{array}\right], \nonumber 
\end{align}

\noindent where $\boldsymbol{G}_2^T = \boldsymbol{G}_{k-1,\ell}^T(j-1:k-1) \hdots \boldsymbol{G}_{j,\ell}^T(j-1:k-1)$. The matrix $\boldsymbol{I} - \boldsymbol{e}_{k-1} \boldsymbol{e}_{k-1}^T$ is the identity matrix where the last diagonal entry is zero. The matrix $\left[\begin{array}{c|c}
\boldsymbol{e}_{j-1} & \boldsymbol{0} \\
\boldsymbol{0}       & \boldsymbol{I} - \boldsymbol{e}_{k-1} \boldsymbol{e}_{k-1}^T
\end{array}\right]$ ensures that the shift affects all diagonal entries of $\boldsymbol{H}$ in the current tile column, but not $\tilde{\boldsymbol{r}}_{\ell}$. The application of the Givens rotations yields $\boldsymbol{R}_{\ell}(1:k-1,j:k-1)$ and a new cross-over column $\tilde{\boldsymbol{s}}_{\ell}$. Figure~\ref{fig:reduction-center-tile-column} illustrates the situation.

The diagonal tile is transformed into a triangular matrix $\boldsymbol{R}_{\ell}(j:k-1,j:k-1)$. 
The resulting triangular system $\boldsymbol{R}_{\ell}(j:k-1,j:k-1) \boldsymbol{x}_{\ell}(j:k-1) = \boldsymbol{b}_{\ell}(j:k-1)$ can be solved with backward substitution for every shift. Its solution $\boldsymbol{x}_{\ell}(j:k-1)$ is used in a linear update. Instead of generating $\boldsymbol{R}_{\ell}(1:j-1,j:k-1)$ explicitly, the Givens rotations can again be applied to the right-hand side. Then the update

\begin{align}\label{eq:update}
\boldsymbol{b}_{\ell}&(1:j-1) \gets \boldsymbol{b}_{\ell}(1:j-1)
-\\
& \begin{bmatrix} 
\boldsymbol{h}(1:j-1,j-1) - \lambda_{\ell} \boldsymbol{e}_{j-1} & \boldsymbol{H}(1:j-1,j:k-2) & \tilde{\boldsymbol{r}}_{\ell}(1:j-1) \\
\end{bmatrix} \nonumber \\
&\hspace{8cm}\left(\boldsymbol{G}_2^T 
\begin{bmatrix}
0 \\
\boldsymbol{x}_{\ell}(j:k-1)
\end{bmatrix}\right)  \nonumber
\end{align}

\noindent exposes the shared $\boldsymbol{H}(1:j-1,j:k-2)$. If many right-hand sides are computed simultaneously, the computation can benefit from level-3 BLAS. Specifically, when interpreted as a block matrix multiplication, the update comprises a scalar-vector multiplication with $\boldsymbol{h}(1:j-1,j-1) - \lambda_{\ell} \boldsymbol{e}_{j-1}$ (level-1 BLAS), a matrix--matrix multiplication with $\boldsymbol{H}(1:j-1,j:k-2)$ (level-3 BLAS) and a shift-specific column update with $\tilde{\boldsymbol{r}}_{\ell}(1:j-1)$ (level-1 BLAS).

The column range $j:k-1$ of the center tile is intentionally left open. It is certainly a possibility to have multiple center tile columns. In this case, a realization can iterate over several center tile column from right to left until the leftmost tile column is reached.

\emph{Leftmost tile column:} It remains to transform the top-left diagonal tile of $\boldsymbol{H}$ to triangular form by applying the Givens rotations $\boldsymbol{G}_{j-1,\ell}^T \hdots \boldsymbol{G}_{2,\ell}^T$ yielding
\begin{align*}
\left(\begin{bmatrix}
\boldsymbol{H}(1:j-1,1:j-2) & \tilde{\boldsymbol{s}}_{\ell}
\end{bmatrix} - \lambda_{\ell}(\boldsymbol{I} - \boldsymbol{e}_{j-1} \boldsymbol{e}_{j-1}^T) \right) \boldsymbol{G}_{3}^T =
\boldsymbol{R}_{\ell}(1:j-1,1:j-1),
\end{align*}
where $\boldsymbol{G}_{3}^T = \boldsymbol{G}_{j-1,\ell}^T(1:j-1,1:j-1) \hdots \boldsymbol{G}_{2,\ell}^T(1:j-1,1:j-1)$. The solution of the triangular system $\boldsymbol{R}_{\ell}(1:j-1,1:j-1) \boldsymbol{x}_{\ell}(1:j-1) = \boldsymbol{b}_{\ell}(1:j-1)$ with backward substitution finalizes the backward substitution phase. 

\textbf{Remark.} The computational efficiency of \eqref{eq:update} hinges on the availability of the cross-over column $\tilde{\boldsymbol{r}}_{\ell}$. If $\tilde{\boldsymbol{r}}_{\ell}$ is not stored, the right side of \eqref{eq:update} is
\begin{align*}
&\boldsymbol{b}_{\ell}(1:j-1) - \underbrace{
\boldsymbol{F}
\begin{bmatrix} 
\boldsymbol{H}(1:j-1,j-1:n) \\
\boldsymbol{H}(j-1:n,j-1:n) - \lambda_{\ell} \boldsymbol{I}
\end{bmatrix} \boldsymbol{G}^T_4}_{
\begin{bmatrix}
\tilde{\boldsymbol{s}}_{\ell} & \boldsymbol{R}_{\ell}(1:j-1,j:n)
\end{bmatrix}
} \begin{bmatrix}
0 \\
\boldsymbol{x}_{\ell}(j:k-1) \\
\boldsymbol{0}_{n-k+1}
\end{bmatrix},
\end{align*}
where $\boldsymbol{G}_4^T = \boldsymbol{G}_{n,\ell}^T(j-1:n,j-1:n) \hdots \boldsymbol{G}_{j,\ell}^T(j-1:n,j-1:n)$ and $\boldsymbol{F} = \begin{bmatrix}
\boldsymbol{I}_{j-1 \times j-1} & \boldsymbol{0}_{j-1 \times n-j+1}
\end{bmatrix}$ extracts the top $j-1$ rows and ensures matching dimensions with $\boldsymbol{b}_{\ell}(1:j-1)$.
If the Givens rotations are applied to the right vector encompassing $\boldsymbol{x}_{\ell}(j:k-1)$, fill-in is generated and the vector becomes dense. Hence, no matter if the Givens rotations are applied to $\boldsymbol{H}$ or to $\boldsymbol{x}_{\ell}$, the computation depends on the $n$-th column of $\boldsymbol{H} - \lambda_{\ell}\boldsymbol{I}$. Storing the cross-over column $\tilde{\boldsymbol{r}}$ effectively prunes the computation of the triangular factor. If the cross-over columns are stored, the reduction and the backward substitution phase can be separated. 
The next section introduces a sequential tiled algorithm which will be parallelized with tasks in the section after.

\subsection{A Tiled Algorithm for Solving Shifted Hessenberg Systems}\label{sec:tiled-algo}

This section uses the ideas presented in the previous section to derive a tiled algorithm for the simultaneous solution of shifted Hessenberg systems. The Hessenberg matrix $\boldsymbol{H} \in \mathbb{R}^{n \times n}$ is partitioned into a grid of $N \times N$ tiles. To simplify the presentation of the algorithms, we assume that all tiles are sized $b_{\rm{r}} \times b_{\rm{r}}$ and that $n = N b_{\rm{r}}$. The shifts are processed in batches of size $b_{\rm{c}}$. We assume that the total number of shifts $m$ is an integer multiple of $b_{\rm{c}}$, i.e., $m = M b_{\rm{c}}$. These two partitionings induce a partitioning of the right-hand sides $\boldsymbol{B} \in \mathbb{R}^{n \times m}$ into a grid of $N \times M$ tiles, where all tiles are sized $b_{\rm{r}} \times b_{\rm{c}}$.

The algorithm is designed for task parallelism and is split into three phases. The \emph{reduction phase} computes the Givens rotations that transform the shifted Hessenberg matrix into a triangular factor. Furthermore, it records the cross-over columns that allow computing linear updates in the backward substitution  as in \eqref{eq:update}. The \emph{backward substitution phase} solves the triangular systems. The \emph{backtransform phase} transforms the computed solutions into solutions to the original shifted Hessenberg systems.

\paragraph{Reduction phase}

\begin{algorithm}[h!t]
\SetKwProg{Fn}{function}{}{end}
\caption{Reduction of a shifted Hessenberg tile to triangular form} 
  \label{alg:tiled-reduction1}
  \KwData{$\boldsymbol{H} \in \mathbb{R}^{n \times n-1}$ and cross-over columns $\tilde{\boldsymbol{R}} = [\tilde{\boldsymbol{r}}_1, \hdots, \tilde{\boldsymbol{r}}_m] \in \mathbb{R}^{n \times m}$ such that $\begin{bmatrix}
  \boldsymbol{H} & \tilde{\boldsymbol{r}}_{\ell}
  \end{bmatrix} \in \mathbb{R}^{n \times n}$ is upper Hessenberg, column $\boldsymbol{h}^{\rm{left}} \in \mathbb{R}^n$, shifts $\boldsymbol{\Lambda} = [
  \lambda_1, \hdots, \lambda_m ]^T \in \mathbb{R}^{m}$}
  \KwResult{Components $\boldsymbol{C}=[c(j,\ell)]\in \mathbb{R}^{n \times m}$ and $\boldsymbol{S}=[c(j,\ell)]\in \mathbb{R}^{n \times m}$ of Givens rotations whose application yields an upper triangular $\boldsymbol{R}_{\ell}$ through
  \begin{footnotesize}
  \begin{align}\label{eq:diagonal-tile}
  \left(\left[\begin{array}{c|cc}
\boldsymbol{h}^{\rm{left}} & \boldsymbol{H} & \tilde{\boldsymbol{r}}_{\ell} + \lambda_{\ell}\boldsymbol{e}_n
\end{array}\right] - \left[\begin{array}{c|c}
\boldsymbol{0} & \lambda_{\ell}\boldsymbol{I}_n
\end{array}\right]\right) &\left[\begin{array}{ccc}
\boldsymbol{I}_{n-1} & \boldsymbol{0} & \boldsymbol{0} \\
\boldsymbol{0}       & c(n,\ell)      & -s(n, \ell)    \\
\boldsymbol{0}       & s(n,\ell)      & c(n, \ell)    \\
\end{array}\right] \hspace{3cm}\nonumber \\
\left[\begin{array}{cccc}
\boldsymbol{I}_{n-2} & \boldsymbol{0} & \boldsymbol{0}& \boldsymbol{0}\\
\boldsymbol{0}       & c(n-1,\ell)      & -s(n-1, \ell)   & 0 \\
\boldsymbol{0}       & s(n-1,\ell)      & c(n-1, \ell)    & 0\\
\boldsymbol{0}       & 0              & 0             & 1\\
\end{array}\right]\hdots &\left[\begin{array}{ccc}
c(1,\ell)      & -s(1, \ell)    & \boldsymbol{0} \\
s(1,\ell)      & c(1, \ell)     & \boldsymbol{0} \\
\boldsymbol{0} & \boldsymbol{0} & \boldsymbol{I}_{n-1}\\
\end{array}\right] \nonumber \\= \left[\begin{array}{c|c}
\boldsymbol{0} & \boldsymbol{R}_{\ell}
\end{array}\right].
\end{align}
\end{footnotesize}
}

\Fn{\upshape \textsc{ReduceDiag}($\boldsymbol{H}, \boldsymbol{h}^{\rm{left}}, \boldsymbol{\Lambda}, \tilde{\boldsymbol{R}}$)}{  
  \For{$\ell \gets 1:m$}{
    $\boldsymbol{v} \gets \tilde{\boldsymbol{r}}_{\ell}$\;
    \For{$k \gets n:-1:2$}{
  		
        Determine and record a Givens rotation such that $\begin{bmatrix}
        h(k,k-1) & v(k)
        \end{bmatrix} \begin{bmatrix}
        c(k,{\ell})  & -s(k,{\ell}) \\
        s(k,{\ell})  &  c(k,{\ell})
        \end{bmatrix} = \begin{bmatrix}
        0 & \star
        \end{bmatrix}$\;
  		
  		$\boldsymbol{v}(1:k-2) \gets c(k,{\ell}) \boldsymbol{h}(1:k-2,k-1) + s(k,{\ell}) \boldsymbol{v}(1:k-2)$\;
  		
  		$v(k-1) \gets c(k,{\ell}) \left(h(k-1,k-1) - \lambda_{\ell} \right) + s(k,{\ell}) v(k-1)$\;
	}
  } 

  \If{$\boldsymbol{h}^{\rm{left}} \neq \boldsymbol{0}$}{
        
        Determine and record a Givens rotation such that $\begin{bmatrix}
        h^{\rm{left}}(1) & v(1)
        \end{bmatrix} \begin{bmatrix}
        c(1,{\ell})  & -s(1,{\ell}) \\
        s(1,{\ell})  &  c(1,{\ell})
        \end{bmatrix} = \begin{bmatrix}
        0 & \star
        \end{bmatrix}$\;
  }
  \Else{$c(1,{\ell}) \gets 1; s(1,{\ell}) \gets 0;$ \tcp{Identity matrix (padding)}} 
  
\Return $\boldsymbol{C}, \boldsymbol{S}$\;
}
\end{algorithm}

\begin{algorithm}[t]
\SetKwProg{Fn}{function}{}{end}
\caption{Reduction of an offdiagonal Hessenberg tile}
  \label{alg:tiled-reduction2}
  \KwData{$\boldsymbol{H} \in \mathbb{R}^{k \times n}$, cross-over columns $\tilde{\boldsymbol{R}}^{\rm{right}} = [\tilde{\boldsymbol{r}}_1^{\rm{right}}, \hdots, \tilde{\boldsymbol{r}}_m^{\rm{right}}] \in \mathbb{R}^{k \times m}$, Givens rotations computed by \textsc{ReduceDiag} and stored as $\boldsymbol{C}=[c(j,\ell)] \in \mathbb{R}^{n \times m}$ and $\boldsymbol{S} = [s(j,\ell)] \in \mathbb{R}^{n \times m}$, shifts $\boldsymbol{\Lambda} = [
  \lambda_1, \hdots, \lambda_m ]^T \in \mathbb{R}^{m}$.}
  \KwResult{Cross-over columns $\tilde{\boldsymbol{R}}^{\rm{left}} = [\tilde{\boldsymbol{r}}_1^{\rm{left}}, \hdots, \tilde{\boldsymbol{r}}_m^{\rm{left}}] \in \mathbb{R}^{k \times m}$
  satisfying 
  \begin{footnotesize}
$\left[\begin{array}{c|cc}
\boldsymbol{h}(1:k,1) - \lambda_{\ell} \boldsymbol{e}_k & \boldsymbol{H}(1:k,2:n) & \tilde{\boldsymbol{r}}_{\ell}^{\rm{right}}
\end{array}\right] \left[\begin{array}{ccc}
\boldsymbol{I}_{n-1} & \boldsymbol{0} & \boldsymbol{0} \\
\boldsymbol{0}       & c(n,\ell)      & -s(n, \ell)    \\
\boldsymbol{0}       & s(n,\ell)      & c(n, \ell)    \\
\end{array}\right]$\\
\hspace{3cm}$\left[\begin{array}{cccc}
\boldsymbol{I}_{n-2} & \boldsymbol{0} & \boldsymbol{0}& \boldsymbol{0}\\
\boldsymbol{0}       & c(n-1,\ell)      & -s(n-1, \ell)   & 0 \\
\boldsymbol{0}       & s(n-1,\ell)      & c(n-1, \ell)    & 0\\
\boldsymbol{0}       & 0              & 0             & 1\\
\end{array}\right]\hdots \left[\begin{array}{ccc}
c(1,\ell)      & -s(1, \ell)    & \boldsymbol{0} \\
s(1,\ell)      & c(1, \ell)     & \boldsymbol{0} \\
\boldsymbol{0} & \boldsymbol{0} & \boldsymbol{I}_{n-1}\\
\end{array}\right] =\left[\begin{array}{c|c}
\tilde{\boldsymbol{r}}^{\rm{left}}_{\ell} & \star
\end{array}\right] $.
  \end{footnotesize}
  }

\Fn{\upshape \textsc{ReduceOffdiag}($\boldsymbol{H}, \tilde{\boldsymbol{R}}^{\rm{right}}, \boldsymbol{\Lambda}$)}{  
  \For{$\ell \gets 1:m$}{
    $\boldsymbol{v} \gets \tilde{\boldsymbol{r}}_{\ell}^{\rm{right}}$\;
    \For{$j \gets n:-1:2$}{
  		$\boldsymbol{v}(1:k) \gets c(j,{\ell}) \boldsymbol{h}(1:k,j) + s(j,{\ell}) \boldsymbol{v}(1:k)$\;
	}
  $\tilde{\boldsymbol{r}}_{\ell}^{\rm{left}}(1:k-1) \gets c(1,{\ell}) \boldsymbol{h}(1:k-1,1) + s(1,{\ell})\boldsymbol{v}(1:k-1) $\;
  $\tilde{r}_{\ell}^{\rm{left}}(k) \gets c(1,{\ell}) (h(k,1) - \lambda_{\ell}) + s(1,{\ell}) v(k) $\;
  }

\Return $[\tilde{\boldsymbol{r}}_{1}^{\rm{left}}, \hdots, \tilde{\boldsymbol{r}}_{m}^{\rm{left}}]$\;
}
\end{algorithm}

The reduction phase computes and records the Givens rotations and cross-over columns required for the backward substitution phase. The computation proceeds tile column by tile column. The transformation of a tile column is split into \textsc{ReduceDiag} and \textsc{ReduceOffdiag}. \textsc{ReduceDiag} addresses the second block row of \eqref{eq:center}; \textsc{ReduceOffdiag} concerns the first block row of \eqref{eq:center}.

The kernel \textsc{ReduceDiag} computes the Givens rotations necessary to transform a diagonal tile to triangular form. 
Algorithm~\ref{alg:tiled-reduction1} lists the details for a batch of shifts. It closely resembles Algorithm~\ref{alg:henry}, but omits the backward substitution and adds the cross-over columns. Note that in Algorithm~\ref{alg:tiled-reduction1} \eqref{eq:diagonal-tile} the matrix $\begin{bmatrix}
\boldsymbol{h}^{\rm{left}} & \boldsymbol{H} & \tilde{\boldsymbol{r}}_{\ell}
\end{bmatrix}$ is $n$-by-$(n+1)$ and the Givens rotations are $(n+1)$-by-$(n+1)$ due to the cross-over columns. The case distinction (lines 8--11) aims at supporting the computation of the top-left diagonal tile. For a top-left diagonal tile, the column $\boldsymbol{h}^{\rm{left}}$ does not physically exist. The driver routine \textsc{TiledReduce} discussed further down sets $\boldsymbol{h}^{\rm{left}}$ to a zero column. In this case the computation of the Givens rotation (line 11) is meaningless. The first column of the storage matrices $\boldsymbol{C}$ and $\boldsymbol{S}$ of the Givens rotations is never read and can be viewed as padding. The flop count of Algorithm~\ref{alg:tiled-reduction1} is $1.5 n^2  + \mathcal{O}(n)$ per shift. 

The kernel \textsc{ReduceOffdiag} concerns the first block row of \eqref{eq:center} and is realized in Algorithm~\ref{alg:tiled-reduction2}. It applies the Givens rotations computed in \textsc{ReduceDiag} to an offdiagonal tile and records the left cross-over column. Algorithm~\ref{alg:tiled-reduction2} requires $3nk + \mathcal{O}(k)$ flops per shift if the tile being processed is $k$-by-$n$.

The partitioning of $\boldsymbol{H}$ into tiles and the kernels \textsc{ReduceDiag} and \textsc{ReduceOffdiag} can be combined into a tiled algorithm for the reduction phase. In a tiled algorithm, the row range covered by the first block row of \eqref{eq:center} is split into smaller tile rows. Since only the tiles directly above the diagonal tiles are affected by the shift, a case distinction is necessary. Given the row index range $\mathcal{I} = i:i+b_{\rm r}-1$, the tiled algorithm listed in Algorithm~\ref{alg:tiled-reduction} computes
\begin{align}\label{eq:tiledred}
&\hspace{-.8cm}\left[\begin{array}{c|cc}
\boldsymbol{h}(\mathcal{I},j-1)- \boldsymbol{\beta} & \boldsymbol{H}(\mathcal{I}, j:j+b_{\rm r}-2) & \tilde{\boldsymbol{r}}_{\ell}(\mathcal{I}, j_{\rm{tile}})
\end{array}\right] \nonumber\\
&\left[\begin{array}{ccc}
\boldsymbol{I}_{b_{\rm r}-1} & \boldsymbol{0} & \boldsymbol{0} \\
\boldsymbol{0}       & c(j+b_{\rm r}-1,\ell)      & -s(j+b_{\rm r}-1,\ell)    \\
\boldsymbol{0}       & s(j+b_{\rm r}-1,\ell)      & c(j+b_{\rm r}-1,\ell)    \\
\end{array}\right]  \nonumber\\
&\left[\begin{array}{cccc}
\boldsymbol{I}_{b_{\rm r}-2} & \boldsymbol{0} & \boldsymbol{0} & \boldsymbol{0}\\
\boldsymbol{0}       & c(j+b_{\rm r}-2,\ell)      & -s(j+b_{\rm r}-2,\ell)  &0  \\
\boldsymbol{0}       & s(j+b_{\rm r}-2,\ell)      & c(j+b_{\rm r}-2,\ell)   &0 \\
0 & 0 & 0 & 1
\end{array}\right] \hdots \\
&\left[\begin{array}{ccc}
c(j,\ell)      & -s(j,\ell)    & \boldsymbol{0}      \\
s(j,\ell)      & c(j,\ell)    & \boldsymbol{0}\\
\boldsymbol{0} & \boldsymbol{0} & \boldsymbol{I}_{b_{\rm r}-1}\\
\end{array}\right] =
\left[\begin{array}{c|c}
\tilde{\boldsymbol{r}}_{\ell}(\mathcal{I}, j_{\rm{tile}}-1) & \star
\end{array}\right]
\end{align}
where $N \geq j_{\rm{tile}} > i/b_{\rm r} \geq 1 $ and $\boldsymbol{\beta} = \lambda_{\ell} \boldsymbol{e}_j$ if $j = i + b_{\rm r}$ (superdiagonal tile affected by shift) or $\boldsymbol{\beta} = \boldsymbol{0}$ otherwise (far-from-diagonal tile unaffected by shift). Line 17 of Algorithm~\ref{alg:tiled-reduction} handles the tiles directly above the diagonal tiles. Line 19, by contrast, handles the remaining far-from-diagonal tiles that are unaffected by the shift. The shift is disabled by setting the corresponding parameter to zero.

\begin{algorithm}
\SetKwProg{Fn}{function}{}{end}
\caption{Tiled reduction of a shifted Hessenberg matrix to triangular form and recording of the applied Givens rotations and relevant cross-over columns}
  \label{alg:tiled-reduction}
  \KwData{Hessenberg matrix $\boldsymbol{H} \in \mathbb{R}^{n \times n}$, tile size $b_{\rm{r}}$ with $n = N b_{\rm{r}}$, $N \geq 2$, shifts $\boldsymbol{\Lambda} = [\lambda_1, \hdots, \lambda_m]^T \in \mathbb{R}^m$, tile size $b_{\rm{c}}$ with $m = M b_{\rm{c}}$}
  \KwResult{Givens rotations $\boldsymbol{G}_{j,\ell}^T$ as in \eqref{eq:givens} and stored as $\boldsymbol{C} = [c(j,\ell)]\in \mathbb{R}^{n \times m}$ and $\boldsymbol{S} = [s(j, \ell)] \in \mathbb{R}^{n \times m} $ such that $(\boldsymbol{H} - \lambda_{\ell}\boldsymbol{I}) \boldsymbol{G}_{n,\ell}^T \hdots \boldsymbol{G}_{2, \ell}^T = \boldsymbol{R}_{\ell}$, cross-over columns $\tilde{\boldsymbol{R}}_{\ell} \in \mathbb{R}^{n \times N}$   satisfying \eqref{eq:tiledred}
}

  \Fn{\upshape \textsc{TiledReduce}($\boldsymbol{H}, \boldsymbol{\Lambda}$)}{

   \For{$\ell \gets 1:m$}{

     $\tilde{\boldsymbol{r}}_{\ell}(:,N) \gets \boldsymbol{h}(:,n)$; $\tilde{r}_{\ell}(n,N) \gets \tilde{r}_{\ell}(n,N) - \lambda_{\ell}$\;
   }

  \For{$\ell \gets 1:b_{\rm{c}}:m$}{
    Set the tile column index $j_{\rm{tile}} \gets N$\;
    $j_{\rm{tile}}^{-} \gets j_{\rm{tile}} - 1$\;
    $\mathcal{L} \gets \ell: \ell + b_{\rm{c}}-1 $\tcp*{Index range of batch of shifts}

  	\For{$j \gets n-b_{\rm{r}}+1:-b_{\rm{r}}:b_{\rm{r}}+1$}{

    	$\mathcal{J} \gets j:j + b_{\rm r} - 1$ \tcp*{Index range of tile column}
    	
    	$\mathcal{J}^{-} \gets j-1:j + b_{\rm r} - 2$ \tcp*{Index range shifted by 1}
  	
        Partition $\begin{bmatrix}
\, 
\smash{\underbrace{%
  \begin{matrix}
  \boldsymbol{h}^{\rm{left}}
  \end{matrix}}_{b_{\rm{r}} \times 1}%
}\,
&
\smash{\underbrace{%
  \begin{matrix}
  \boldsymbol{H}^{\rm{diag}}
  \end{matrix}}_{b_{\rm{r}} \times (b_{\rm{r}}-1)}%
}\,
\end{bmatrix}
\vphantom{%
\underbrace{%
  \begin{matrix}
  H^{\rm{diag}}
  \end{matrix}}_{\boldsymbol{b}_{\rm{r}} \times b_{\rm{r}}}%
} \gets \boldsymbol{H}(\mathcal{J},\mathcal{J}^{-})$\;
  	
    Pack $\tilde{\boldsymbol{R}}^{\rm{right}} \gets \left[\tilde{\boldsymbol{r}}_{\ell}(\mathcal{J}, j_{\rm{tile}}), \tilde{\boldsymbol{r}}_{\ell}(\mathcal{J},j_{\rm{tile}}), ..., \tilde{\boldsymbol{r}}_{\ell + b_{\rm{c}}-1}(\mathcal{J},j_{\rm{tile}}) \right]$\;
  	
  		\mbox{$\boldsymbol{C}(\mathcal{J}, \mathcal{L}), \boldsymbol{S}(\mathcal{J}, \mathcal{L}) \gets \textsc{ReduceDiag}(\boldsymbol{H}^{\rm{diag}},\boldsymbol{h}^{\rm{left}},\boldsymbol{\Lambda}(\mathcal{L}), \tilde{\boldsymbol{R}}^{\rm{right}})$}\;

  		\For{$i \gets j-b_{\rm{r}}:-b_{\rm{r}}:1$}{
  			$\mathcal{I}\gets i:i+b_{\rm r}-1$ \tcp*{Index range of tile row}	
  		
          \mbox{Pack $\tilde{\boldsymbol{R}}^{\rm{right}} \gets \left[\tilde{\boldsymbol{r}}_{\ell}(\mathcal{I},j_{\rm{tile}}), \tilde{\boldsymbol{r}}_{\ell +1}(\mathcal{I},j_{\rm{tile}}), \hdots, \tilde{\boldsymbol{r}}_{\ell + b_{\rm{r}}-1}(\mathcal{I},j_{\rm{tile}}) \right]$}\;  		
  		
            \If{$i + b_{\rm{r}} = j$}{
                $[\tilde{\boldsymbol{r}}_{\ell}(\mathcal{I},j_{\rm{tile}}^{-}),\tilde{\boldsymbol{r}}_{\ell +1}(\mathcal{I},j_{\rm{tile}}^{-}),\hdots, \tilde{\boldsymbol{r}}_{\ell + b_{\rm{r}}-1}(\mathcal{I},j_{\rm{tile}}^{-})] \gets\textsc{ReduceOffdiag}(\boldsymbol{H}(\mathcal{I}, \mathcal{J}^{-}), \tilde{\boldsymbol{R}}^{\rm{right}}, \boldsymbol{\Lambda}(\mathcal{L}))$\;            
            }
            \Else{
  		      $[\tilde{\boldsymbol{r}}_{\ell}(\mathcal{I},j_{\rm{tile}}^{-}),\tilde{\boldsymbol{r}}_{\ell +1}(\mathcal{I},j_{\rm{tile}}^{-}),\hdots, \tilde{\boldsymbol{r}}_{\ell + b_{\rm{r}}-1}(\mathcal{I},j_{\rm{tile}}^{-})] \gets\textsc{ReduceOffdiag}(\boldsymbol{H}(\mathcal{I}, \mathcal{J}^{-}), \tilde{\boldsymbol{R}}^{\rm{right}}, \boldsymbol{0})$\;            
            }
        }
  		
  		$j_{\rm{tile}} \gets j_{\rm{tile}} - 1$\;
  	}

	$\mathcal{J} \gets 1:b_{\rm r}$ \tcp*{Index range of top-left corner}
    \mbox{Pack $\tilde{\boldsymbol{R}}^{\rm{right}} \gets \left[\tilde{\boldsymbol{r}}_{\ell}(\mathcal{J},1), \tilde{\boldsymbol{r}}_{\ell+1}(\mathcal{J},1), \hdots, \tilde{\boldsymbol{r}}_{\ell + b_{\rm{c}}-1}(\mathcal{J},1) \right]$}\;
\mbox{$\boldsymbol{C}(\mathcal{J}, \mathcal{L}), \boldsymbol{S}(\mathcal{J}, \mathcal{L}) \gets \textsc{ReduceDiag}(\boldsymbol{H}(\mathcal{J},1:b_{\rm r}-1), \boldsymbol{0}, \boldsymbol{\Lambda}(\mathcal{L}), \tilde{\boldsymbol{R}}^{\rm{right}})$}\;

  }   
   \Return $\boldsymbol{C},\boldsymbol{S},(\tilde{\boldsymbol{R}}_1, \hdots, \tilde{\boldsymbol{R}}_m)$\;
  }
\end{algorithm}

\paragraph{Backward substitution phase}
The backward substitution phase computes the solution to $\boldsymbol{R}_{\ell}\boldsymbol{x}_{\ell} = \boldsymbol{b}_{\ell}$ for every $\ell$. The computation follows a standard pattern of tiled backward substitution. It iterates over tiles of the solution from bottom to top. For each iteration a new tile of the solution is computed with a small backward substitution algorithm. The small backward substitution is realized by Algorithm~\ref{alg:solve} \textsc{Solve}. Then the readily computed part of the solution is used in a tile update of above-lying tiles, realized by Algorithm~\ref{alg:tile-update} \textsc{Update}. The combination of these two kernels yields the tiled backward substitution algorithm \textsc{TiledSolve} listed in Algorithm~\ref{alg:tiled-solve}. In the following we present the details of the algorithms.

\begin{algorithm}[t]
\SetKwProg{Fn}{function}{}{end}
\caption{Small backward substitution}
  \label{alg:solve}
  \KwData{$\boldsymbol{H} \in \mathbb{R}^{n \times n-1}$ and cross-over columns $\tilde{\boldsymbol{R}} =[\tilde{\boldsymbol{r}}_{1}, \hdots, \tilde{\boldsymbol{r}}_{m}] \in \mathbb{R}^{n \times m}$ such that $\begin{bmatrix}
  \boldsymbol{H} & \tilde{\boldsymbol{r}}_{\ell}
  \end{bmatrix} \in \mathbb{R}^{n \times n}$ is upper Hessenberg, column $\boldsymbol{h}^{\rm{left}}$, $\boldsymbol{X} \in \mathbb{R}^{n \times m}$, shifts $\boldsymbol{\Lambda} = [\lambda_1, \hdots, \lambda_m]^T\in \mathbb{R}^m$, $\boldsymbol{B} \in \mathbb{R}^{n \times m}$, Givens rotations computed by \textsc{TiledReduce} and stored as $\boldsymbol{C} = [c(j,{\ell})] \in \mathbb{R}^{n \times m}$ and $\boldsymbol{S} = [s(j,{\ell})] \in \mathbb{R}^{n \times m}$.}
  \KwResult{$\boldsymbol{X} \in \mathbb{R}^{n\times m}$ satisfying $\boldsymbol{R}_{\ell} \boldsymbol{x}(:,\ell) = \boldsymbol{b}(:,\ell)$ where $\boldsymbol{R}_{\ell}$ is given by \eqref{eq:diagonal-tile}.}

  \Fn{\upshape \textsc{Solve}($\boldsymbol{H}, \boldsymbol{h}^{\rm{left}}, \boldsymbol{\Lambda}, \tilde{\boldsymbol{R}}, \boldsymbol{C}, \boldsymbol{S}, \boldsymbol{B}$)}{
      $\boldsymbol{X} \gets \boldsymbol{B}$\;
      \For{$\ell \gets 1:m$}{
      	$\boldsymbol{v} \gets \tilde{\boldsymbol{r}}_{\ell}$\;
      	Compute $\boldsymbol{x}(2:n,\ell)$ using lines 2--9 of Algorithm~\ref{alg:henry}\;
      	\If{$\boldsymbol{h}^{\rm{left}} \neq \boldsymbol{0}$}{
          $v(1) \gets v(1) c(1,\ell) - h^{\rm{left}}(1) s(1,\ell)$\;
        }
        $x(1,\ell) \gets x(1,\ell)/v(1)$\;
      }

	\Return $\boldsymbol{X}$\;
  }
\end{algorithm}

\textsc{Solve} realizes the small backward substitution. 
Since our algorithm splits the reduction phase and the backward substitution phase, the relevant part of the triangular factor has to be recomputed. Algorithm~\ref{alg:solve} gives the details. 
Analogously to Algorithm~\ref{alg:tiled-reduction1} \textsc{ReduceDiag}, the top-left diagonal tile of $\boldsymbol{H}-\lambda_{\ell}\boldsymbol{I}$ does not have a physical column $\boldsymbol{h}^{\rm{left}}$ to its left. In that case, the driver routine \textsc{TiledSolve} discussed below sets $\boldsymbol{h}^{\rm{left}}$ to a  zero column and thereby skips the computation of the left cross-over column (line 7). The flop count of Algorithm~\ref{alg:solve} is $3.5n^2 + \mathcal{O}(n)$.

Algorithm~\ref{alg:tile-update} \textsc{Update} realizes \eqref{eq:update}. The application of the Givens rotations in \eqref{eq:update} to the right transforms
\begin{displaymath}
\boldsymbol{G}_{k-1,\ell}^T(j-1:k-1) \hdots \boldsymbol{G}_{j,\ell}^T(j-1:k-1) 
\begin{bmatrix}
0 \\
\boldsymbol{x}_{\ell}(j:k-1)
\end{bmatrix} =
\begin{bmatrix}
\rho \\
\boldsymbol{z}_{\ell}
\end{bmatrix},
\end{displaymath}
which is computed in lines 3 and 9--13. The block matrix multiplication \eqref{eq:update} with the left block yields the scalar-vector multiplication $\boldsymbol{b}_{\ell} \gets \boldsymbol{b}_{\ell} - \rho (\boldsymbol{h}(1:j-1,j-1) - \lambda_{\ell} \boldsymbol{e}_{j-1}) $ (line 4--5). The multiplication with the center block $\boldsymbol{H}(1:j-1,j:k-2)$ is executed in line 14. When many vectors $\boldsymbol{z}_{\ell}$ are computed simultaneously, this operation corresponds to a matrix--matrix multiplication. Consequently, it can benefit from an efficient implementation of \textsc{DGEMM} available in an optimized BLAS library~\cite{BLAS,BLASupdated}. The multiplication with the right block $\tilde{\boldsymbol{r}}_{\ell}$ is realized with the vector-scalar multiplication (line 16). The total flop count of Algorithm~\ref{alg:tile-update} is $2mnk + \mathcal{O}((m+k)n)$. When several right-hand sides are computed simultaneously, the matrix--matrix multiplication in line 14 dominates the computation. 

The partitioning into tiles combined with the routines \textsc{Solve} and \textsc{Update} leads to the tiled backward substitution algorithm \textsc{TiledSolve} listed in Algorithm~\ref{alg:tiled-solve}. Offdiagonal tile column updates are split into smaller tile row updates. Only the tiles directly above diagonal tiles are affected by the shift (line 16). Far-from-diagonal tile updates, by contrast, are unaffected by the shifts (line 18). The computation of $\boldsymbol{z}_{\ell}$ is repeated for each tile row. Based on numerical experiments, these additional flops are negligible compared to the gain from task parallelism, which will be introduced in the next section.

\begin{algorithm}[t]
\SetKwProg{Fn}{function}{}{end}
\caption{Linear Tile Update}
  \label{alg:tile-update}
  \KwData{$\boldsymbol{H} \in \mathbb{R}^{m \times k}$, $\boldsymbol{X} \in \mathbb{R}^{k \times n}$, $\boldsymbol{B} \in \mathbb{R}^{m \times n}$,
  cross-over columns $\tilde{\boldsymbol{R}} =[\tilde{\boldsymbol{r}}_{1}, \hdots, \tilde{\boldsymbol{r}}_{n}] \in \mathbb{R}^{m \times n}$, 
  $\boldsymbol{\Lambda}=[\lambda_1, \hdots, \lambda_{n}]^T \in \mathbb{R}^n$,
  Givens rotations computed by \textsc{TiledReduce} and stored as $\boldsymbol{C} = [c(j,{\ell})] \in \mathbb{R}^{k \times n}$ and $\boldsymbol{S} = [s(j,{\ell})] \in \mathbb{R}^{k \times n}$ }
  \KwResult{\eqref{eq:update} for $\ell = 1, \hdots, n$}

  \Fn{\upshape \textsc{Update}($\boldsymbol{H}, \tilde{\boldsymbol{R}}, \boldsymbol{X}, \boldsymbol{\Lambda}, \boldsymbol{C}, \boldsymbol{S}, \boldsymbol{B}$)}{ 
    \For{$\ell \gets 1:n$}{
        $\rho \gets s(1,{\ell}) x(1,\ell)$\;
        $\boldsymbol{b}(1:m,\ell) \gets \boldsymbol{b}(1:m,\ell) + \rho \boldsymbol{h}(1:m,1)$\;
  	 	$b(m,\ell) \gets b(m,\ell) - \lambda_{\ell} \rho$\;
	}

    Copy $\boldsymbol{Z} \gets \boldsymbol{X}$\;
    \For{$\ell \gets 1:n$}{
       $z(1,\ell) \gets c_{\ell}(1) z(1,\ell)$\;
       \For{$j \gets 2:k$}{
       		$\tau_1 \gets z(j-1,\ell)$\;
       		$\tau_2 \gets z(j,\ell)$\;
       		$z(j-1,\ell) \gets c(j,{\ell}) \tau_1 - s(j,{\ell}) \tau_2$\;
       		$z(j,\ell) \gets s(j,{\ell}) \tau_1 + c(j,{\ell}) \tau_2$\;
		}
	}
	
	$\boldsymbol{B} \gets \boldsymbol{B} - \boldsymbol{H}(1:m,2:k) \boldsymbol{Z}(1:k-1,1:n)$ \tcp*{DGEMM}
	
	\For{$\ell \gets 1:n$}{
		$\boldsymbol{b}(1:m,\ell) \gets \boldsymbol{b}(1:m,\ell) - \tilde{\boldsymbol{r}}_{\ell} z(k,\ell)$\;
	}

	\Return $\boldsymbol{B}$\;
  }
\end{algorithm}

\begin{algorithm}
\SetKwProg{Fn}{function}{}{end}
\caption{Tiled backward substitution and backtransform}
  \label{alg:tiled-solve}
  \KwData{Hessenberg matrix $\boldsymbol{H} \in \mathbb{R}^{n \times n}$, tile size $b_{\rm{r}}$ with $n = N b_{\rm{r}}$  and $N \geq 2$, shifts $\boldsymbol{\Lambda} = [\lambda_1, \hdots, \lambda_m]^T \in \mathbb{R}^m$, $\boldsymbol{B} \in \mathbb{R}^{n \times m}$, tile size $b_{\rm{c}}$ with $m = M b_{\rm{c}}$, Givens rotations $\boldsymbol{G}_{j,\ell}^T$ stored as $\boldsymbol{C} \in \mathbb{R}^{n \times m}$ and $\boldsymbol{S}\in \mathbb{R}^{n \times m}$ and cross-over columns $\tilde{\boldsymbol{R}}_{\ell} \in \mathbb{R}^{n \times N}$ computed with Algorithm~\ref{alg:tiled-reduction} for $\ell = 1, \hdots, m$.}
\KwResult{Solution $\boldsymbol{X} \in \mathbb{R}^{n \times m}$ to $\boldsymbol{H} \boldsymbol{X} = \boldsymbol{B} \operatorname{diag}(\lambda_1, \hdots, \lambda_m)$}

  \Fn{\upshape \textsc{TiledSolve}($\boldsymbol{H}, \boldsymbol{\Lambda}, \boldsymbol{C}, \boldsymbol{S}, (\tilde{\boldsymbol{R}}_1, \hdots, \tilde{\boldsymbol{R}}_m), \boldsymbol{B}$)}{
   $\boldsymbol{X} \gets \boldsymbol{B}$\;
   \For{$\ell \gets 1:b_{\rm{c}}:m$}{
    $j_{\rm{tile}} \gets N$\tcp*{Index of tile column}
    $\mathcal{L} \gets \ell: \ell + b_{\rm{c}}-1 $\tcp*{Index range of batch of shifts}
    
    \For{$j \gets n-b_{\rm{r}}+1:-b_{\rm{r}}:1$}{
        $\mathcal{J} \gets j:j+b_{\rm{r}}-1$ \tcp*{Index range of tile column}
        $\mathcal{J}^{-} \gets j-1:j+b_{\rm{r}}-2$ \tcp*{Index range shifted by 1}

Partition $\begin{bmatrix}
\, 
\smash{\underbrace{%
  \begin{matrix}
  \boldsymbol{h}^{\rm{left}}
  \end{matrix}}_{b_{\rm{r}} \times 1}%
}\,
&
\smash{\underbrace{%
  \begin{matrix}
  \boldsymbol{H}^{\rm{diag}}
  \end{matrix}}_{b_{\rm{r}} \times (b_{\rm{r}}-1)}%
}\,
\end{bmatrix}
\vphantom{%
\underbrace{%
  \begin{matrix}
  H^{\rm{diag}}
  \end{matrix}}_{\boldsymbol{b}_{\rm{r}} \times b_{\rm{r}}}%
} \gets \boldsymbol{H}(\mathcal{J},\mathcal{J}^{-})$\;

        Pack $\tilde{\boldsymbol{R}}^{\rm{right}} \gets \left[\tilde{\boldsymbol{r}}_{\ell}(\mathcal{J},j_{\rm{tile}}), \hdots, \tilde{\boldsymbol{r}}_{\ell + b_{\rm{c}}-1}(\mathcal{J},j_{\rm{tile}}) \right]$\; 
        
        $\boldsymbol{X}(\mathcal{J}, \mathcal{L}) \gets \textsc{Solve}(\boldsymbol{H}^{\rm{diag}},\boldsymbol{h}^{\rm{left}}, \tilde{\boldsymbol{R}}^{\rm{right}}, \boldsymbol{\Lambda}(\mathcal{L}), \boldsymbol{C}(\mathcal{J},\mathcal{L}), \boldsymbol{S}(\mathcal{J},\mathcal{L}), \boldsymbol{X}(\mathcal{J}, \mathcal{L}))$\;

        \For{$i \gets j-b_{\rm{r}}:-b_{\rm{r}}:1$}{
            $\mathcal{I} \gets i:i+b_{\rm{r}}-1$ \tcp*{Index range of tile row}            
              
           Pack~\mbox{$\tilde{\boldsymbol{R}}^{\rm{right}} \gets \left[\tilde{\boldsymbol{r}}_{\ell}(\mathcal{I},j_{\rm{tile}}), \tilde{\boldsymbol{r}}_{\ell +1}(\mathcal{I},j_{\rm{tile}}), \hdots, \tilde{\boldsymbol{r}}_{\ell + b_{\rm{c}}-1}(\mathcal{I},j_{\rm{tile}}) \right]$}\;

          \If{$i + b_{\rm{r}} = j$}{
              $\boldsymbol{X}(\mathcal{I}, \mathcal{L}) \gets \textsc{Update}(\boldsymbol{H}(\mathcal{I}, \mathcal{J}^-), \tilde{\boldsymbol{R}}^{\rm{right}}, \boldsymbol{X}(\mathcal{J}, \mathcal{L}), \newline \hphantom{aaaaaaaaaaaaaaaa.} \boldsymbol{\Lambda}(\mathcal{L}), \boldsymbol{C}(\mathcal{J},\mathcal{L}), \boldsymbol{S}(\mathcal{J},\mathcal{L}), \boldsymbol{X}(\mathcal{I}, \mathcal{L}))$\;           
          }
          \Else{
              $\boldsymbol{X}(\mathcal{I}, \mathcal{L}) \gets \textsc{Update}(\boldsymbol{H}(\mathcal{I}, \mathcal{J}^-), \tilde{\boldsymbol{R}}^{\rm{right}}, \boldsymbol{X}(\mathcal{J}, \mathcal{L}), \newline \hphantom{aaaaaaaaaaaaaaaa.}\boldsymbol{0}, \boldsymbol{C}(\mathcal{J},\mathcal{L}), \boldsymbol{S}(\mathcal{J},\mathcal{L}), \boldsymbol{X}(\mathcal{I}, \mathcal{L}))$\;  
          }
        }
        $j_{\rm{tile}} \gets j_{\rm{tile}} - 1$\;
    }
    $\mathcal{J} \gets 1:b_{\rm r}$\;
    Pack~$\tilde{\boldsymbol{R}}^{\rm{right}} \gets \left[\tilde{\boldsymbol{r}}_{\ell}(\mathcal{J},j_{\rm{tile}}), \hdots, \tilde{\boldsymbol{r}}_{\ell + b_{\rm{c}}-1}(\mathcal{J},j_{\rm{tile}}) \right]$\; 
    $\boldsymbol{X}(\mathcal{J}, \mathcal{L}) \gets \textsc{Solve}(\boldsymbol{H}(\mathcal{J}, 1:b_{\rm r}-1),\boldsymbol{0}, \boldsymbol{\Lambda}(\mathcal{L}), \newline \hphantom{aaaaaaaaaaaaaaa.}\tilde{\boldsymbol{R}}^{\rm{right}}, \boldsymbol{C}(\mathcal{J},\mathcal{L}), \boldsymbol{S}(\mathcal{J},\mathcal{L}), \boldsymbol{X}(\mathcal{J}, \mathcal{L}))$\;
    $\boldsymbol{X}(:,\mathcal{L}) \gets \textsc{Backtransform}(\boldsymbol{C}(:,\mathcal{L}), \boldsymbol{S}(:,\mathcal{L}), \boldsymbol{X}(:,\mathcal{L}))$\;
  }
  \Return $\boldsymbol{X}$\;
  }
\end{algorithm}

\paragraph{Backtransform phase}

The solution $\boldsymbol{x}_{\ell}$ corresponding to a shift $\lambda_\ell$ is backtransformed
\begin{equation}\label{eq:backtransform}
\boldsymbol{y}_{\ell} \gets \boldsymbol{G}_{n,\ell}^T \left(\hdots  \left(\boldsymbol{G}_{2,\ell}^T \boldsymbol{x}_{\ell}\right)\right).
\end{equation}
The components of the Givens rotations $\boldsymbol{G}_{j,\ell}^T$ \eqref{eq:givens} have been recorded as $\boldsymbol{C}=[c(j,\ell)] \in \mathbb{R}^{n\times m}$ and $\boldsymbol{S}=[s(j,\ell)]\in \mathbb{R}^{n\times m}$ in the reduction phase. 
Analogously to the batched processing in the reduction and the backward substitution phase, a routine $\boldsymbol{Y}(1:n,\mathcal{L}) \gets \textsc{Backtransform}(\boldsymbol{C}(1:n,\mathcal{L}), \boldsymbol{S}(1:n,\mathcal{L}), \boldsymbol{X}(1:n,\mathcal{L}))$ realizes \eqref{eq:backtransform} for a batch of vectors covering the contiguous index range $\mathcal{L}$. A possible implementation is a column-by-column backtransform with the lines 11--14 of Algorithm~\ref{alg:henry}. The flop count for the backtransform is $\mathcal{O}(n)$ per shift. Algorithm~\ref{alg:tiled-solve} \textsc{TiledSolve} merges the backward substitution phase and the backtransform phase.

\begin{algorithm}
\SetKwProg{Fn}{function}{}{end}
\caption{Solving $(\boldsymbol{H} - \lambda_{\ell} \boldsymbol{I})\boldsymbol{x}_{\ell} = \boldsymbol{b}_{\ell}$ simultaneously in a tiled fashion}
  \label{alg:nonrobust-hsrq3}

    $\boldsymbol{C}, \boldsymbol{S}, (\tilde{\boldsymbol{R}}_1, \hdots, \tilde{\boldsymbol{R}}_{m}) \gets \textsc{TiledReduce}(\boldsymbol{H}, \boldsymbol{\Lambda})$\;
    $\boldsymbol{X} \gets \textsc{TiledSolve}(\boldsymbol{H}, \boldsymbol{\Lambda}, \boldsymbol{C}, \boldsymbol{S}, (\tilde{\boldsymbol{R}}_1, \hdots, \tilde{\boldsymbol{R}}_{m}), \boldsymbol{B})$\;
\end{algorithm}

The successive execution of \textsc{TiledReduce} and \textsc{TiledSolve} yields a tiled solver for the simultaneous solution of shifted Hessenberg systems. This solver is listed in Algorithm~\ref{alg:nonrobust-hsrq3} and can be viewed as an extension to Algorithm~\ref{alg:henry}.

\subsection{Task Parallelism}\label{sec:task-parallelism}
This section introduces task parallelism to the kernels introduced in the previous section. Since Algorithm~\ref{alg:nonrobust-hsrq3} calls \textsc{TiledReduce} and \textsc{TiledSolve} successively, our parallel implementation separates these two kernels with a synchronization point. Then it remains to taskify \textsc{TiledReduce} and \textsc{TiledSolve}.

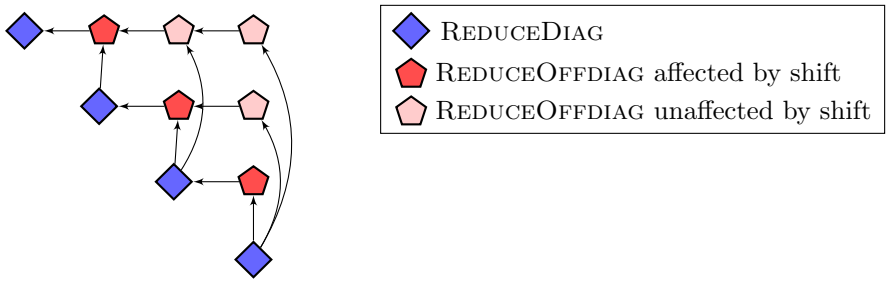
\begin{figure}
\hspace{.2cm}
\begin{tikzpicture}[
  node distance = .6cm,
  reducediag/.style = {draw, regular polygon, regular polygon sides=4, shape border rotate=45, thick,fill=blue!60},
  reduceoffdiag/.style = {draw, regular polygon, regular polygon sides=5, thick,fill=red!20},
  reduceoffdiag2/.style = {draw, regular polygon,regular polygon sides=5, thick,fill=red!70},
  arrow/.style = {-latex'}]

\node[reducediag] (reducediag) {};
\node[reduceoffdiag2, above = of reducediag] (reduceoffdiag1) {};
\node[reduceoffdiag, above = of reduceoffdiag1] (reduceoffdiag2) {};
\node[reduceoffdiag, above = of reduceoffdiag2] (reduceoffdiag3) {};
\node[reduceoffdiag2, left = of reduceoffdiag2] (reduceoffdiag4) {};
\node[reducediag, left = of reduceoffdiag1] (reducediag1) {};
\node[reducediag, left = of reduceoffdiag4] (reducediag2) {};
\node[reduceoffdiag, left = of reduceoffdiag3] (reduceoffdiag5) {};
\node[reduceoffdiag2, left = of reduceoffdiag5] (reduceoffdiag6) {};
\node[reducediag, left = of reduceoffdiag6] (reducediag3) {};

\draw[arrow](reducediag) to (reduceoffdiag1);
\draw[arrow](reducediag1) to (reduceoffdiag4);
\draw[arrow](reducediag) to [bend right=30] (reduceoffdiag2);
\draw[arrow](reducediag) to [bend right=30] (reduceoffdiag3);
\draw[arrow](reducediag1) to [bend right=30] (reduceoffdiag5);
\draw[arrow](reducediag2) to (reduceoffdiag6);

\draw[arrow](reduceoffdiag1) to (reducediag1);
\draw[arrow](reduceoffdiag4) to (reducediag2);
\draw[arrow](reduceoffdiag6) to (reducediag3);
\draw[arrow](reduceoffdiag3) to (reduceoffdiag5);
\draw[arrow](reduceoffdiag5) to (reduceoffdiag6);
\draw[arrow](reduceoffdiag2) to (reduceoffdiag4);

\node[reducediag, label=right:\textsc{ReduceDiag}, right = of reduceoffdiag5, xshift=2cm](legend1) {};
\node[reduceoffdiag2, label=right:\textsc{ReduceOffdiag} affected by shift, below = of legend1, yshift = .5cm] {};
\node[reduceoffdiag, label=right:\textsc{ReduceOffdiag} unaffected by shift, below = of legend1, yshift = -0.0cm] {};
\node[draw, below = of legend1, yshift=1.2cm, xshift=2.9cm ,minimum width=6.6cm, minimum height=1.7cm]() {};
\end{tikzpicture}
\caption{Task graph constructed for a single batch of shifts and a tile column count of $n/b_{\rm{r}} = 4$ in Algorithm~\ref{alg:tiled-reduction}. The task arrangement is based on what tile of $\boldsymbol{H}$ is processed.}
\label{fig:task-graph-tiled-reduction}
\end{figure}

Algorithm~\ref{alg:tiled-reduction} \textsc{TiledReduce} can be taskified by defining each function call as a task. The data dependences between tasks are as follows. A task \textsc{ReduceDiag} processing the tile $(j,j)$ of $\boldsymbol{H}$ has outgoing dependences to \textsc{ReduceOffdiag} tasks processing above-lying tiles $(i,j)$ in the same tile column of $\boldsymbol{H}$, $i = 1, \hdots, j-1$. A \textsc{ReduceOffdiag} task on $(i,j)$ has an outgoing dependence to the task processing the left-lying tile $(i,j-1)$. This is either a \textsc{ReduceDiag} tasks ($j-1 = i$) or another \textsc{ReduceOffdiag} task ($j-1 > i$). Figure~\ref{fig:task-graph-tiled-reduction} illustrates the task graph for one batch of shifts, i.e., one iteration of line 4 in Algorithm~\ref{alg:tiled-reduction}. Batches of shifts are independent of each other. In other words, there is one task graph per iteration of the loop in line 4.

Algorithm~\ref{alg:tiled-solve} \textsc{TiledSolve} is taskified using three task types. First, every call to \textsc{Update} yields a task. Second, all calls to \textsc{Solve} except for the one processing the top-left tile $(1,1)$ of $\boldsymbol{H}$ correspond to a task. Third, the remaining \textsc{Solve} on $(1,1)$ and the \textsc{Backtransform} are merged into one task. A \textsc{Solve} task on $(j,j)$ has outgoing dependences to \textsc{Update} tasks $(i,j)$, $i = 1, \hdots, j-1$. Before a \textsc{Solve} on the tile $(j,j)$ can be executed, all updates $(i,j)$, $j > i$ must have been completed. Figure~\ref{fig:task-graph-tiled-solve} shows the task graph for one iteration of line 3 in Algorithm~\ref{alg:tiled-solve}. The computation of \textsc{Solve} on $(1,1)$ and the \textsc{Backtransform} are merged because of two reasons. First, it reduces the amount of dependences that have to be handled by the runtime system. Second, unless merged, the computationally cheap backtransform incurs significant scheduling overhead. Analogously to the task-parallel execution of \textsc{TiledReduce}, batches of shifts do not have any dependences.

\begin{figure}
\centering
\begin{tikzpicture}[
  node distance = .6cm,
  backsolve/.style = {draw, circle, thick,fill=teal!60},
  update/.style = {draw, regular polygon, regular polygon sides=4, thick},
  update2/.style = {draw, regular polygon,regular polygon sides=4, thick,fill=black!20},
  transform/.style = {draw, regular polygon,regular polygon sides=6, thick,fill=yellow!80},
  arrow/.style = {-latex'}]

\node[backsolve] (backsolve) {};
\node[update2, above = of backsolve] (update1) {};
\node[update, above = of update1] (update2) {};
\node[update, above = of update2] (update3) {};
\node[update2, left = of update2] (update4) {};
\node[backsolve, left = of update1] (solve1) {};
\node[backsolve, left = of update4] (solve2) {};
\node[update, left = of update3] (update5) {};
\node[update2, left = of update5] (update6) {};
\node[transform, left = of update6] (transform) {};

\draw[arrow](backsolve) to (update1);
\draw[arrow](solve1) to (update4);
\draw[arrow](backsolve) to [bend right=30] (update2);
\draw[arrow](backsolve) to [bend right=30] (update3);
\draw[arrow](solve1) to [bend right=30] (update5);
\draw[arrow](solve2) to (update6);

\draw[arrow](update1) to (solve1);
\draw[arrow](update4) to (solve2);
\draw[arrow](update6) to (transform);
\draw[arrow](update3) to [bend right=35] (transform);
\draw[arrow](update5) to [bend right=21] (transform);
\draw[arrow](update2) to [bend right=25] (solve2);

\node[backsolve, label=right:\textsc{Solve}, above = of update5, xshift=-3.cm, yshift=2cm](legend1) {};
\node[update2, label=right:\textsc{Update} affected by shift, below = of legend1, yshift = .45cm] {};
\node[update, label=right:\textsc{Update} unaffected by shift, below = of legend1, yshift = 0cm] {};
\node[transform, label=right:Merged \textsc{Solve} and \textsc{Backtransform}, below = of legend1, yshift = -.45cm] {};

\node[draw, below = of legend1, yshift=1.1cm, xshift=2.9cm ,minimum width=6.8cm, minimum height=2.cm]() {};
\end{tikzpicture}
\caption{Task graph constructed for a single batch of shifts and a tile column count of $n/b_{\rm{r}} = 4$ in Algorithm~\ref{alg:tiled-solve}. The task arrangement is based on what tile of $\boldsymbol{H}$ is processed.}
\label{fig:task-graph-tiled-solve}
\end{figure}

\section{Robust Computation of Eigenvectors by Inverse Iteration}\label{sec:inverse-iteration}

This section extends Algorithm~\ref{alg:nonrobust-hsrq3} for solving shifted Hessenberg systems to the computation of eigenvectors. The algorithm relies on the ability to solve triangular systems. The solution of triangular systems is known to be prone to overflow. This is particularly true if shifted Hessenberg systems are solved as part of an inverse iteration algorithm. In that case $\boldsymbol{H}-\lambda \boldsymbol{I}$ can be expected to be ill-conditioned if $\lambda$ is close to a true eigenvalue of $\boldsymbol{H}$. By adding overflow protection, Section~\ref{sec:overflow} renders Algorithm~\ref{alg:nonrobust-hsrq3} robust while preserving the tiled structure. This yields the robust, tiled shifted Hessenberg system solver \textsc{DHSRQ3}. Section~\ref{sec:starting-vector} concerns the convergence criterion and the choice of the starting vector. Together with the robust, tiled shifted Hessenberg solver, this results in a robust tiled inverse iteration algorithm \textsc{DHSRQ3IN}. Section~\ref{sec:complex-shifts} discusses the modifications necessary to support complex shifts.

\subsection{Ill-conditioned Systems and Overflow Protection}\label{sec:overflow}
Inverse iteration assumes that a good approximation $\lambda$ to a true eigenvalue of $\boldsymbol{H}$ is available. In that case $\lambda$ is an exact eigenvalue of a nearby matrix $\boldsymbol{H} + \boldsymbol{E}$ where $\| \boldsymbol{E} \| = \mathcal{O}(\epsilon)$ is of the order of the machine precision. The matrix $\boldsymbol{H} + \boldsymbol{E} - \lambda \boldsymbol{I}$ is exactly singular and, hence, $\boldsymbol{H} - \lambda \boldsymbol{I}$ is close-to-singular. Once $\boldsymbol{H} - \lambda \boldsymbol{I}$ has been factored, the backward substitution with the corresponding triangular factor solves an ill-conditioned system. The computed solution can be so large that the representational range is exceeded. To eliminate the possiblity of such a floating-point overflow, implementations of inverse iteration~\cite[p. 435]{peters1971calculation} and shifted Hessenberg system solvers~\cite{henry1994techreport,henry1995parallel} introduce a scaling factor $\gamma \in (0,1]$ and solve the scaled triangular linear system $\boldsymbol{R} \boldsymbol{x} = \gamma \boldsymbol{b}$ for the scaled solution $\gamma^{-1}\boldsymbol{x}$. By virtue of $\gamma$, the current representation of the solution can be rescaled such that overflow is avoided. This renders the solution process robust. A robust solver for this scaled triangular system is, for example, \textsc{DLATRS}~\cite{anderson1991robust} available in LAPACK 3.9.0.

An extension to the solution of scaled triangular linear systems with $n$ right-hand sides has been introduced by Kjelgaard Mikkelsen and Karlsson~\cite{mikkelsen2017blocked} and concerns $\boldsymbol{R} \boldsymbol{X} = \boldsymbol{B} \operatorname{diag}(\gamma_1, \hdots, \gamma_n)$. Each right-hand-side is associated with a scaling factor  $\gamma_k \in (0,1]$. If $\boldsymbol{B}$ is overwritten with the solution $\boldsymbol{X}$ in a standard (non-robust) tiled backward substitution algorithm, tile updates read $\boldsymbol{X}_i \gets \boldsymbol{X}_{i} - \boldsymbol{R}_{ij} \boldsymbol{X}_j, i<j$. These tile updates are rendered robust through segment-wise scaling factors. There is one scaling factor per column per tile. Then robust tile updates are $\boldsymbol{X}_{i} \operatorname{diag}(\delta^{-1}_{1}, \hdots, \delta^{-1}_{n}) \gets \boldsymbol{X}_{i} \operatorname{diag}(\beta^{-1}_{1}, \hdots, \beta^{-1}_{n}) - \boldsymbol{R}_{ij} (\boldsymbol{X}_{j} \operatorname{diag}(\alpha^{-1}_{1}, \hdots, \alpha^{-1}_{n}))$. Overflow is avoided by bounding the maximum possible growth in each update, computing suited scaling factors $\delta^{-1}_{1}, \hdots, \delta^{-1}_{n}$ and, if necessary, rescaling the current representation of the solution prior to the matrix--matrix multiplication. This preserves the level-3 BLAS potential within a tile update. Since tile updates during the tiled backward substitution can require different scalings, the final tiled representation of the solution can be inconsistently scaled. A consistent scaling is computed by reducing the segment-wise scaling factors to the global scaling factor $\gamma_k$ for each column. Then all column segments are rescaled with respect to $\gamma_k$. This approach supports task parallelism as demonstrated by Kjelgaard Mikkelsen et al.~\cite{kjelgaard2019parallel} and is used to render the algorithms presented in this paper robust.

This paper adopts Henry's assumption that $\boldsymbol{H} - \lambda \boldsymbol{I}$ can be factored into R and Q without encountering overflow. Then a robust tiled algorithm for solving shifted Hessenberg systems requires (a) a robust routine for the solution of small shifted Hessenberg systems, (b) a robust tile update and (c) a robust backtransform. The combination of these three robust kernels yields a robust version of \textsc{TiledReduce}. The robust routines are marked with the prefix \textsc{R} to easily distinguish between the non-robust and the robust version.

\paragraph{Robust solution of small shifted Hessenberg systems} A robust counterpart of Algorithm~\ref{alg:solve} \textsc{Solve} requires a small robust backward substitution routine solving $\boldsymbol{R}_{\ell} \boldsymbol{x}_{\ell} = \gamma_{\ell} \boldsymbol{b}_{\ell}$, $\gamma_{\ell} \in (0,1]$. If the (small) $\boldsymbol{R}_{\ell}$ is generated explicitly, a possible realization is a call to the LAPACK routine \textsc{DLATRS}. This approach is realized by Algorithm~\ref{alg:robust-solve} \textsc{RSolve}. \textsc{DLATRS} returns $\boldsymbol{x}_{\ell}$ and $\gamma_{\ell}$ representing $\gamma_{\ell}^{-1}\boldsymbol{x}_{\ell}$. The total scaling of the computed solution is computed in line 13 by multiplying the input scaling factor $\beta_{\ell}$ and $\gamma_{\ell}$. Together, \textsc{RSolve} returns the scaled vector $(\gamma_{\ell} \beta_{\ell})^{-1}\boldsymbol{x}_{\ell}$.

\begin{algorithm}[t]
\SetKwProg{Fn}{function}{}{end}
\caption{Small robust backward substitution}
  \label{alg:robust-solve}
\KwData{$\boldsymbol{H} \in \mathbb{R}^{n \times n-1}$ and cross-over columns $\tilde{\boldsymbol{R}} =[\tilde{\boldsymbol{r}}_{1}, \hdots, \tilde{\boldsymbol{r}}_{m}] \in \mathbb{R}^{n \times m}$ such that $\begin{bmatrix}
  \boldsymbol{H} & \tilde{\boldsymbol{r}}_{\ell}
  \end{bmatrix} \in \mathbb{R}^{n \times n}$ is upper Hessenberg, column $\boldsymbol{h}^{\rm{left}}$, $\boldsymbol{B} \in \mathbb{R}^{n \times m}$ and scaling factors $\boldsymbol{\beta}=[\beta_1, \hdots, \beta_m]=(0,1]^m$ representing $\boldsymbol{B}\operatorname{diag}(\beta_1, \hdots, \beta_m)$, shifts $\boldsymbol{\Lambda} = [\lambda_1, \hdots, \lambda_m]^T\in \mathbb{R}^m$, Givens rotations computed by \textsc{TiledReduce} and stored as $\boldsymbol{C} = [c(j,{\ell})] \in \mathbb{R}^{n \times m}$ and $\boldsymbol{S} = [s(j,{\ell})] \in \mathbb{R}^{n \times m}$.}
  \KwResult{$\boldsymbol{X} \in \mathbb{R}^{n\times m}$ and $\boldsymbol{\alpha}=[\alpha_1, \hdots, \alpha_m] \in (0,1]^m$ satisfying $\boldsymbol{R}_{\ell} \boldsymbol{x}(:,\ell) = \alpha_{\ell}\boldsymbol{b}(:,\ell)$ where $\boldsymbol{R}_{\ell}$ is given by \eqref{eq:diagonal-tile}.}

  \Fn{\upshape \textsc{RSolve}($\boldsymbol{H}, \boldsymbol{h}^{\rm{left}}, \tilde{\boldsymbol{R}}, \boldsymbol{\Lambda}, \boldsymbol{C}, \boldsymbol{S}, \boldsymbol{\beta}, \boldsymbol{B}$)}{
    $\boldsymbol{X} \gets \boldsymbol{B}$\;
    \For{$\ell \gets 1:m$}{
      $\gamma_{\ell} \gets 1$\;
      $\boldsymbol{R}_{\ell} \gets \boldsymbol{0}_{n \times n}$\;
      $\boldsymbol{r}_{\ell}(:,n) \gets \tilde{\boldsymbol{r}}_{\ell}$\;
      \For{$k \gets n:-1:2$}{
         $\boldsymbol{t}_{\ell} \gets \boldsymbol{H}(1:k,k-1)$; $t_{\ell}(k-1) \gets t(k-1) - \lambda_{\ell}$\;
      	 $\boldsymbol{R}_{\ell}(1:k,k-1:k) \gets \begin{bmatrix}
      	 \boldsymbol{t}_{\ell}(1:k) & \boldsymbol{r}_{\ell}(1:k,k)
      	 \end{bmatrix}\begin{bmatrix}
      	 c(k,\ell) \\
      	 s(k,\ell)
      	 \end{bmatrix}$\;
      }
      \If{$\boldsymbol{h}^{\rm{left}} \neq \boldsymbol{0}$}{
        $r_{\ell}(1,\ell) \gets r_{\ell}(1,\ell) c(1,\ell) - h^{\rm{left}}(1) s(1,\ell)$\;
      }
      Solve robustly $\boldsymbol{R}_{\ell}\boldsymbol{x}(:,\ell) = \gamma_{\ell} \boldsymbol{b}(:,\ell)$\tcp*{DLATRS}
      $\alpha_{\ell} \gets \gamma_{\ell} \beta_{\ell}$\;
    }
	\Return $\boldsymbol{\alpha}$, $\boldsymbol{X}$\;
  }
\end{algorithm}

\paragraph{Robust tile update}

\begin{algorithm}[th!]
\SetKwProg{Fn}{function}{}{end}
\caption{Robust Linear Tile Update}
  \label{alg:robust-tile-update}
  \KwData{$\boldsymbol{H} \in \mathbb{R}^{m \times k}$, $\boldsymbol{X} \in \mathbb{R}^{k \times n}$, $\boldsymbol{B} \in \mathbb{R}^{m \times n}$,
  cross-over columns $\tilde{\boldsymbol{R}} =[\tilde{\boldsymbol{r}}_{1}, \hdots, \tilde{\boldsymbol{r}}_{n}] \in \mathbb{R}^{m \times n}$, 
  $\boldsymbol{\Lambda}=[\lambda_1, \hdots, \lambda_{n}]^T \in \mathbb{R}^n$,
  Givens rotations computed by \textsc{TiledReduce} and stored as $\boldsymbol{C} = [c(j,{\ell})] \in \mathbb{R}^{k \times n}$ and $\boldsymbol{S} = [s(j,{\ell})] \in \mathbb{R}^{k \times n}$ }
  \KwResult{\eqref{eq:robust-update} for $\ell = 1, \hdots, n$}

  \Fn{\upshape \textsc{RUpdate}($\boldsymbol{H}, \tilde{\boldsymbol{R}}, \boldsymbol{\alpha}, \boldsymbol{X}, \boldsymbol{\Lambda}, \boldsymbol{C}, \boldsymbol{S}, \boldsymbol{\beta}, \boldsymbol{B}$)}{
    Allocate $\boldsymbol{\delta} = [\delta_1, \hdots, \delta_n] \in (0,1]^n$\;
    \For{$\ell \gets 1:n$}{
        $\boldsymbol{t} \gets \boldsymbol{h}(1:m,1)$; $t(m) \gets t(m) - \lambda_{\ell}$\;
        $\gamma_{\ell} \gets \min\{\alpha_{\ell}, \beta_{\ell}\}$\;
        $\xi_{\ell} \gets \textsc{ProtectUpdate}\left( \left\| \frac{\gamma_{\ell}}{\beta_{\ell}} \boldsymbol{b}(1:m,\ell) \right\|_{\infty}, \left\| \boldsymbol{t} \right\|_{\infty}, \left| \frac{\gamma_{\ell}}{\alpha_{\ell}} s(1, \ell) x(1,\ell) \right|\right)$\;
        $\delta_{\ell} \gets \xi_{\ell} \gamma_{\ell}$\;
  	 	$\boldsymbol{b}(1:m,\ell) \gets \frac{\delta_{\ell}}{\beta_{\ell}} \boldsymbol{b}(1:m,\ell) - \boldsymbol{t} \left(\frac{\delta_{\ell}}{\beta_{\ell}}  s(1,{\ell}) x(1,\ell) \right)$\;
	}

    Copy $\boldsymbol{Z} \gets \boldsymbol{X} \operatorname{diag}\left(\frac{\delta_1}{\alpha_1}, \hdots, \frac{\delta_n}{\alpha_n}\right)$\;
    Apply Givens rotations to $\boldsymbol{Z}$ as in lines 7--13 of Algorithm~\ref{alg:tile-update}\;
	
	\For{$\ell \gets 1:n$}{
	    $\xi_{\ell} \gets \textsc{ProtectUpdate}(\|\boldsymbol{b}(1:m,\ell) \|_{\infty}, \left\| \boldsymbol{H} \right\|_{\infty}, \| \boldsymbol{z}(1:k-1, \ell) \|_{\infty})$\;
        $\delta_\ell \gets \delta_\ell \xi_{\ell}$\;
	    $\boldsymbol{b}(1:m,\ell) \gets \xi_{\ell} \boldsymbol{b}(1:m,\ell)$; $\boldsymbol{z}(1:k,\ell) \gets \xi_{\ell} \boldsymbol{z}(1:k,\ell)$\;
	}
	
	$\boldsymbol{B} \gets \boldsymbol{B} - \boldsymbol{H}(1:m,2:k) \boldsymbol{Z}(1:k-1,1:n)$ \tcp*{DGEMM}
	
	\For{$\ell \gets 1:n$}{
	    $\xi_{\ell} \gets \textsc{ProtectUpdate}(\|\boldsymbol{b}(1:m,\ell) \|_{\infty}, \| \tilde{\boldsymbol{r}}_{\ell} \|_{\infty}, |z(k, \ell)| )$\;
	    $\delta_\ell \gets \delta_\ell \xi_{\ell}$\;
		$\boldsymbol{b}(1:m,\ell) \gets \xi_{\ell}\boldsymbol{b}(1:m,\ell) - \tilde{\boldsymbol{r}}_{\ell} \left( \xi_{\ell}z(k,\ell) \right)$\;
	}

	\Return $\boldsymbol{\delta}$, $\boldsymbol{B}$\;
  }
\end{algorithm}

The robust version of the tile update \eqref{eq:update} is
\begin{align}\label{eq:robust-update}
\delta_{\ell}^{-1} \boldsymbol{b}_{\ell}(1:j-1) \gets \beta_{\ell}^{-1} \boldsymbol{b}_{\ell}(1:j-1) - \hspace{6cm}& \nonumber \\
\begin{bmatrix} 
\boldsymbol{h}(1:j-1,j-1) - \lambda_{\ell} \boldsymbol{e}_{j-1} & \boldsymbol{H}(1:j-1,j:k-2) & \tilde{\boldsymbol{r}}_{\ell}(1:j-1) \\
\end{bmatrix} \qquad \quad & \nonumber \\
\left(\boldsymbol{G}_2^T
\begin{bmatrix}
0 \\
\alpha_{\ell}^{-1} \boldsymbol{x}_{\ell}(j:k-1)
\end{bmatrix}
\right)&,
\end{align}
where $\delta_l \in (0,1]$ is chosen such that $\boldsymbol{b}_{\ell}(1:j-1)$ does not exceed the overflow threshold. There are many instantiations of $\delta_l$ and $\boldsymbol{b}_{\ell}(1:j-1)$ that satisfy \eqref{eq:update}. Algorithm~\ref{alg:robust-tile-update} computes one feasible instantiation. Following Kjelgaard Mikkelsen and Karlsson~\cite{mikkelsen2017nlafet}, the right-hand side tile $\boldsymbol{X}$ is associated with a vector of scaling factors $\boldsymbol{\alpha} = [\alpha_1, \hdots, \alpha_n] \in (0,1]^n$ and represents the column-wise scaled matrix $\boldsymbol{X} \operatorname{diag}(\alpha_1^{-1}, \hdots, \alpha_n^{-1})$. Similarly, the tile $\boldsymbol{B}$ is associated with $\boldsymbol{\beta} = [\beta_1, \hdots, \beta_n] \in (0,1]^n$ and represents $\boldsymbol{B} \operatorname{diag}(\beta_1^{-1}, \hdots, \beta_n^{-1})$. To compute a tile update robustly, $\boldsymbol{X}$ and $\boldsymbol{B}$ must be consistently scaled. The consistent scaling factor corresponds to the smaller of the two scaling factors (line 5). The remaining computation requires the overflow-free realization of three linear updates.

Each of the three linear updates is guarded by \textsc{ProtectUpdate} introduced by Kjelgaard Mikkelsen and Karlsson~\cite[Section 2.2]{mikkelsen2017nlafet}. \textsc{ProtectUpdate} receives $\| \boldsymbol{B} \|_\infty$, $\| \boldsymbol{H} \|_\infty$, $\| \boldsymbol{X} \|_\infty$ and computes a scaling factor $\xi \in (0,1]$ such that the linear update $\xi \boldsymbol{B} - \boldsymbol{H} (\xi \boldsymbol{X})$ cannot overflow. Then line 6 computes the column-wise scaling factors required for the first linear update (line 8). The upper bounds are rescaled to account for a consistent scaling of $\boldsymbol{B}$ and $\boldsymbol{X}$. After rescaling $\boldsymbol{X}$ (line 9), $\boldsymbol{B}$ and $\boldsymbol{X}$ are consistently scaled. Line 12 computes the column-wise scaling factors required for the second linear update (line 15). As line 14 applies the computed scaling factors prior to this linear update, the linear update itself can safely be implemented with a call to \textsc{DGEMM}. Line 17 computes the scaling factors necessary for the third linear update (line 19). The scaling of the final output $\boldsymbol{B}$ is $\boldsymbol{\delta}$ and corresponds to the product of all scaling factors.

The application of the Givens rotations in line 10 is not guarded by overflow protection logic, but can result in growth that possibly exceeds the overflow threshold. Since the evaluation of overflow protection logic is expensive compared to the cheap Givens transformations, our software lowers the true overflow threshold $\Omega$ by some safety margin. This safety margin is set to the tile height $b_{\rm r}$, which overestimates the maximum growth possible by Givens transformations within a tile. Working with $\Omega/b_{\rm r} $ is cheap to compute and guarantees that the Givens transformations in line 10 do not trigger overflow.

\paragraph{Robust backtransform}

\begin{algorithm}[th!]
\SetKwProg{Fn}{function}{}{end}
\caption{Consistency scaling, backtransform and normalization of a single eigenvector}
  \label{alg:consistency-scaling}
  \KwData{Givens rotations $\boldsymbol{G}_{j,\ell}^T$ as in \eqref{eq:givens} computed by \textsc{TiledReduce} and stored as $\boldsymbol{C} = [c(j,\ell)] \in \mathbb{R}^{n \times m}$ and $\boldsymbol{S} = [s(j,\ell)] \in \mathbb{R}^{n \times m}$, matrix of scaling factors $\boldsymbol{\alpha} = [\alpha(i,\ell)]\in (0,1]^{N \times m}$, segment-wise scaled $\boldsymbol{X} \in \mathbb{R}^{n \times m}$, tile size $b_{\rm r}$ with $n = N b_{\rm r}$
  }
  \KwResult{Normalized \eqref{eq:robust-backtransform} for $\ell \gets 1, \hdots, m$
  }

  \Fn{\upshape \textsc{RBacktransform}($\boldsymbol{C}, \boldsymbol{S}, \boldsymbol{\alpha}, \boldsymbol{X}$)}{

       \For{$\ell \gets 1:m$}{
         \tcp{Global scaling factor of current column}
         $\alpha_{\min} \gets \min \limits_{1 \leq h \leq N} \{\alpha(h,\ell)\}$\;
         
         $x_{\max} \gets 0$; $t \gets 0$\;
         \For{$i \gets 1:n$}{
           $i_{\rm{tile}} \gets \lceil i/b_{\rm r} \rceil$ \tcp*{Index of current segment}

           \If{$x(i,\ell) \neq 0$}{
             $x(i,\ell) \gets \frac{\alpha_{\min}}{\alpha(i_{\rm{tile},\ell})} x(i,\ell)$ \tcp*{Consistency scaling}
             \If{$x_{\max} < |x(i,\ell)|$}{
               $t \gets 1 + t \left(\frac{x_{\max}}{|x(i,\ell)|}\right)^2$;
               $x_{\max} \gets |x(i,\ell)|$\;
             }
             \Else{
               $t \gets t + \left(\frac{|x(i,\ell)|}{x_{\max}}\right)^2$\;
             }
           }
         }
         $x_{\rm{nrm}} \gets \alpha_{\min}^{-1} x_{\max} \sqrt{t}$ \tcp*{Euclidean norm}
         $\tau_1 \gets  x(1,\ell)/x_{\rm{nrm}}$\;
         \For{$i \gets 2:n$}{
           $\tau_2 \gets x(i,\ell)/x_{\rm{nrm}}$\;
           $y(i-1,\ell) \gets c(i) \tau_1 - s(i,\ell) \tau_2$\;
           $\tau_1 \gets c(i,\ell) \tau_2 + s(i,\ell) \tau_1$\;
         }
     }
     \Return $\boldsymbol{Y}$ \tcp*{$\| \boldsymbol{y}(:,\ell) \|_2 = 1$}

  }
\end{algorithm}

The robust backward substitution returns segment-wise scaled solution vectors. It remains to compute consistently scaled solutions and backtransform these. Recall that the partitioning into tile rows divides the vector evenly into $N$ segments of length $b_{\rm r}$ and that $n = N b_{\rm r}$. The robust counterpart of \eqref{eq:backtransform} is then
\begin{equation}\label{eq:robust-backtransform}
\boldsymbol{y}_{\ell} = \boldsymbol{G}_{n,\ell}^T \left(\hdots \left(\boldsymbol{G}_{2,\ell}^T
\alpha_{\min}^{-1}
\begin{bmatrix}
\frac{\alpha_{\min}}{\alpha_1} \boldsymbol{x}_{\ell}(1:b_{\rm r}) \\
\vdots \\
\frac{\alpha_{\min}}{\alpha_N} \boldsymbol{x}_{\ell}(n-b_{\rm r}+1 : n)
\end{bmatrix}\right)\right).
\end{equation}
The application of the Givens rotations can exceed the overflow threshold. To avoid overflow, Henry~\cite[Algorithm 4]{henry1994techreport} evaluates the maximum possible growth possible during the backtransform. If overflow \emph{can} occur, the entire vector is rescaled prior to the backtransform. If the computation targets eigenvectors, an alternative strategy is possible. Eigenvectors are commonly normalized. The backtransform can be executed safely if the vector is normalized with respect to the Euclidean norm before the backtransform. The consistency scaling, the normalization and the backtransform can be computed in two sweeps over the vector. Algorithm \ref{alg:consistency-scaling} \textsc{RBacktransform} gives the details. The lines 4--12 closely follow the LAPACK routine \textsc{DNRM2}, which computes the Euclidean norm with scaling to avoid overflow. The lines 14--18 simultaneously normalize and backtransform an eigenvector. After the backtransform, the eigenvectors are still normalized.

\begin{algorithm}[h!]
\SetKwProg{Fn}{function}{}{end}
\caption{Robust tiled backward substitution and backtransform}
  \label{alg:robust-tiled-solve}
  \KwData{As in Algorithm~\ref{alg:tiled-solve}.
  }
\KwResult{Solution $\boldsymbol{X} \in \mathbb{R}^{n \times m}$ to $\boldsymbol{H} \boldsymbol{X} = \boldsymbol{B} \operatorname{diag}(\lambda_1, \hdots, \lambda_m)$}

  \Fn{\upshape \textsc{RobustTiledSolve}($\boldsymbol{H}, \boldsymbol{\Lambda}, \boldsymbol{C}, \boldsymbol{S}, (\tilde{\boldsymbol{R}}_1, \hdots, \tilde{\boldsymbol{R}}_m), \boldsymbol{B}$)}{
   $\boldsymbol{\alpha} \gets \textsc{ones}(N,m)$; $\boldsymbol{X} \gets \boldsymbol{B}$\;
   \For{$\ell \gets 1:b_{\rm{c}}:m$}{
    $j_{\rm{tile}} \gets N$\tcp*{Index of current tile column}
    $\mathcal{L} \gets \ell: \ell + b_{\rm{c}}-1 $\tcp*{Index range of batch of shifts}
    
    \For{$j \gets n-b_{\rm{r}}+1:-b_{\rm{r}}:1$}{
        $\mathcal{J} \gets j:j+b_{\rm{r}}-1$; $\mathcal{J}^{-} \gets j-1:j+b_{\rm{r}}-2$

Partition $\begin{bmatrix}
\, 
\smash{\underbrace{%
  \begin{matrix}
  \boldsymbol{h}^{\rm{left}}
  \end{matrix}}_{b_{\rm{r}} \times 1}%
}\,
&
\smash{\underbrace{%
  \begin{matrix}
  \boldsymbol{H}^{\rm{diag}}
  \end{matrix}}_{b_{\rm{r}} \times (b_{\rm{r}}-1)}%
}\,
\end{bmatrix}
\vphantom{%
\underbrace{%
  \begin{matrix}
  H^{\rm{diag}}
  \end{matrix}}_{\boldsymbol{b}_{\rm{r}} \times b_{\rm{r}}}%
} \gets \boldsymbol{H}(\mathcal{J},\mathcal{J}^{-})$\;

        Pack $\tilde{\boldsymbol{R}}^{\rm{right}} \gets \left[\tilde{\boldsymbol{r}}_{\ell}(\mathcal{J},j_{\rm{tile}}), \hdots, \tilde{\boldsymbol{r}}_{\ell + b_{\rm{c}}-1}(\mathcal{J},j_{\rm{tile}}) \right]$\; 
        
        $\boldsymbol{\alpha}(j_{\rm{tile}},\mathcal{L}), \boldsymbol{X}(\mathcal{J}, \mathcal{L}) \gets \textsc{RSolve}(\boldsymbol{H}^{\rm{diag}},\boldsymbol{h}^{\rm{left}}, \tilde{\boldsymbol{R}}^{\rm{right}}, \boldsymbol{\Lambda}(\mathcal{L}),\newline \hphantom{.........} \boldsymbol{C}(\mathcal{J},\mathcal{L}), \boldsymbol{S}(\mathcal{J},\mathcal{L}), \boldsymbol{\alpha}(j_{\rm{tile}},\mathcal{L}), \boldsymbol{X}(\mathcal{J}, \mathcal{L}))$\;

        \For{$i \gets j-b_{\rm{r}}:-b_{\rm{r}}:1$}{
            $\mathcal{I} \gets i:i+b_{\rm{r}}-1$ \tcp*{Index range of tile row}
            $i_{\rm{tile}} \gets (i+b_{\rm r}-1)/b_{\rm r}$\tcp*{Index of current tile row}
              
           \mbox{Pack $\tilde{\boldsymbol{R}}^{\rm{right}} \gets \left[\tilde{\boldsymbol{r}}_{\ell}(\mathcal{I},j_{\rm{tile}}), \tilde{\boldsymbol{r}}_{\ell +1}(\mathcal{I},j_{\rm{tile}}), ..., \tilde{\boldsymbol{r}}_{\ell + b_{\rm{c}}-1}(\mathcal{I},j_{\rm{tile}}) \right]$}\;
                     
          \If{$i + b_{\rm{r}} = j$}{
              $\boldsymbol{\alpha}(i_{\rm{tile}},\mathcal{L}), \boldsymbol{X}(\mathcal{I}, \mathcal{L}) \gets \textsc{RUpdate}(\boldsymbol{H}(\mathcal{I}, \mathcal{J}^-), \tilde{\boldsymbol{R}}^{\rm{right}}, \boldsymbol{\alpha}(j_{\rm{tile}},\mathcal{L}), \boldsymbol{X}(\mathcal{J}, \mathcal{L}), \newline \hphantom{.........} \boldsymbol{\Lambda}(\mathcal{L}), \boldsymbol{C}(\mathcal{J},\mathcal{L}), \boldsymbol{S}(\mathcal{J},\mathcal{L}),\boldsymbol{\alpha}(i_{\rm{tile}},\mathcal{L}), \boldsymbol{X}(\mathcal{I}, \mathcal{L}))$\;
          }
          \Else{
              $\boldsymbol{\alpha}(i_{\rm{tile}},\mathcal{L}), \boldsymbol{X}(\mathcal{I}, \mathcal{L}) \gets \textsc{RUpdate}(\boldsymbol{H}(\mathcal{I}, \mathcal{J}^-), \tilde{\boldsymbol{R}}^{\rm{right}}, \boldsymbol{\alpha}(j_{\rm{tile}},\mathcal{L}), \boldsymbol{X}(\mathcal{J}, \mathcal{L}),  \newline \hphantom{.........}\boldsymbol{0}, \boldsymbol{C}(\mathcal{J},\mathcal{L}), \boldsymbol{S}(\mathcal{J},\mathcal{L}), \boldsymbol{\alpha}(i_{\rm{tile}},\mathcal{L}), \boldsymbol{X}(\mathcal{I}, \mathcal{L}))$\;  
          }
        }
        $j_{\rm{tile}} \gets j_{\rm{tile}} - 1$\;
    }
    $\mathcal{J} \gets 1:b_{\rm r}$\;
    Pack $\tilde{\boldsymbol{R}}^{\rm{right}} \gets \left[\tilde{\boldsymbol{r}}_{\ell}(\mathcal{J},j_{\rm{tile}}), \hdots, \tilde{\boldsymbol{r}}_{\ell + b_{\rm{c}}-1}(\mathcal{J},j_{\rm{tile}}) \right]$\; 
    $\boldsymbol{\alpha}(1,\mathcal{L}), \boldsymbol{X}(\mathcal{J}, \mathcal{L}) \gets \textsc{RSolve}(\boldsymbol{H}(\mathcal{J}, 1:b_{\rm r}-1),\boldsymbol{0}, \boldsymbol{\Lambda}(\mathcal{L}),\newline \hphantom{.........} \tilde{\boldsymbol{R}}^{\rm{right}}, \boldsymbol{C}(\mathcal{J},\mathcal{L}), \boldsymbol{S}(\mathcal{J},\mathcal{L}), \boldsymbol{\alpha}(1,\mathcal{L}), \boldsymbol{X}(\mathcal{J}, \mathcal{L}))$\;
    \mbox{$\boldsymbol{X}(:,\mathcal{L}) \gets \textsc{RBacktransform}(\boldsymbol{C}(:,\mathcal{L}), \boldsymbol{S}(:,\mathcal{L}), \boldsymbol{\alpha}(:,\mathcal{L}), \boldsymbol{X}(:,\mathcal{L}))$}\;
  }
  \Return $\boldsymbol{X}$\;
  }
\end{algorithm}

By replacing all routines with their robust counterparts and adding segment-wise scaling factors, Algorithm~\ref{alg:tiled-solve} \textsc{TiledSolve} can be rendered robust. The resulting algorithm \textsc{RobustTiledSolve} is listed in Algorithm~\ref{alg:robust-tiled-solve}. This leads to \textsc{DHSRQ3} listed in Algorithm~\ref{alg:hsrq3}, which solves shifted Hessenberg systems in a tiled, robust fashion. Note that the first part \textsc{TiledReduce} is untouched. Only the second part \textsc{TiledSolve} is replaced with its robust counterpart. Since the structure is identical to the non-robust version, task parallelism as introduced in Section~\ref{sec:task-parallelism} is valid for \textsc{DHSRQ3} as well.

\begin{algorithm}[t]
\SetKwProg{Fn}{function}{}{end}
\caption{Robust tiled simultaneous solution of shifted Hessenberg systems with real shifts}
  \label{alg:hsrq3}
  \KwData{Hessenberg matrix $\boldsymbol{H} \in \mathbb{R}^{n \times n}$, shifts $\boldsymbol{\Lambda} = [\lambda_1, \hdots, \lambda_m]^T \in \mathbb{R}^m$, $\boldsymbol{B} \in \mathbb{R}^{n \times m}$}
  \KwResult{Solution $\boldsymbol{X} \in \mathbb{R}^{n \times m}$ to $\boldsymbol{H} \boldsymbol{X} = \boldsymbol{B} \operatorname{diag}(\lambda_1, \hdots, \lambda_m)$}

  \Fn{\upshape \textsc{DHSRQ3}($\boldsymbol{H}, \boldsymbol{\Lambda}, \boldsymbol{B}$)}{
    $\boldsymbol{C}, \boldsymbol{S}, (\tilde{\boldsymbol{R}}_1, \hdots, \tilde{\boldsymbol{R}}_{m}) \gets \textsc{TiledReduce}(\boldsymbol{H}, \boldsymbol{\Lambda})$\;
    $\boldsymbol{X} \gets \textsc{RobustTiledSolve}(\boldsymbol{H}, \boldsymbol{\Lambda}, \boldsymbol{C}, \boldsymbol{S}, (\tilde{\boldsymbol{R}}_1, \hdots, \tilde{\boldsymbol{R}}_{m}), \boldsymbol{B})$\;
    \Return $\boldsymbol{X}$\;
  }
\end{algorithm}

\subsection{Starting Vector and Convergence Test}\label{sec:starting-vector}
It is well understood that the choice of the starting vector is crucial for both the convergence and the performance of inverse iteration \eqref{eq:inviteration}. This is particularly true because the residual can increase by doing more than one iteration. Ipsen~\cite[Sections 2.5, 2.6, 6.2]{ipsen1997inverseiteration} presents a comprehensive summary of work by Varah, Wilkinson and Peters on choosing a suited starting vector. Furthermore, an example demonstrating an increasing residual when more than one iteration is computed can be found in Section 5.4 in the same reference.

A standard choice for the starting vector is a scaled vector of ones $\boldsymbol{x}^{(0)} \gets \rho [1, \hdots, 1]^T$, $\rho > 0$~\cite[p. 259]{ipsen1997inverseiteration}. In LAPACK 3.9.0 the inverse iteration routine \textsc{DHSEIN} chooses $\rho = \| \boldsymbol{H} \|_\infty \epsilon$ assuming $\| \boldsymbol{H} \|_\infty > 0$. With this choice a single iteration of \eqref{eq:inviteration} most frequently leads to convergence \cite[p. 264]{ipsen1997inverseiteration}. If not, LAPACK tries other starting vectors orthogonal to previous choices rather than computing more iterations of \eqref{eq:inviteration}. The convergence test is passed if $\|\boldsymbol{x}^{(1)}\|_1 > 0.1/\sqrt{n}$. LAPACK thereby follows Varah's~\cite[p. 786]{varah1968calculation} stopping criterion
$\frac{\| \boldsymbol{x}^{(1)} \|_2}{\| \boldsymbol{x}^{(0)} \|_2} > \frac{1}{c \epsilon},$
where $c$ is a problem-dependent constant and $\boldsymbol{H}$ is assumed to be normalized $\|\boldsymbol{H}\|_2 = 1$. Specifically, since LAPACK does not assume $\|\boldsymbol{H}\|_2 = 1$, we obtain using $\| \boldsymbol{x}^{(0)} \|_1 = n \| \boldsymbol{H} \|_{\infty} \epsilon$, $1/\|\boldsymbol{x} \|_2 \geq  1/\|\boldsymbol{x} \|_1$ and $ \| \boldsymbol{x} \|_2 \geq \|\boldsymbol{x} \|_1 / \sqrt{n}$ 
\begin{displaymath}
\frac{\| \boldsymbol{x}^{(1)} \|_2}{\| \boldsymbol{x}^{(0)} \|_2} \geq \frac{\| \boldsymbol{x}^{(1)} \|_1}{\sqrt{n}\| \boldsymbol{x}^{(0)} \|_1}  = \frac{ \|\boldsymbol{x}^{(1)} \|_1}{\sqrt{n} n \| \boldsymbol{H} \|_\infty \epsilon} > \frac{1}{10 n^2 \| \boldsymbol{H} \|_\infty \epsilon}.
\end{displaymath}

Our inverse iteration solver chooses $\rho = \|\boldsymbol{H} \|_\infty \epsilon$. Since the norm $\|\boldsymbol{x}^{(1)}\|_2$ is readily available after the backward substitution phase due to the consistency scaling, the check $\|\boldsymbol{x}^{(1)}\|_2 > 0.1/\sqrt{n}$ lends itself to a quick convergence test. This convergence test is in line with Varah's stopping criterion where $c = 10n \| \boldsymbol{H} \|_\infty$. The information that an eigenvector has not converged can, for example, be propagated by setting the eigenvector to zero.

The decision on a starting vector and the convergence test combined with the robust backward substitution leads to the inverse iteration routine \textsc{HSRQ3IN}, a routine for computing individual eigenvectors simultaneously by inverse iteration. The core of the routine is the robust, tiled solver for shifted Hessenberg systems introduced as \textsc{DHSRQ3}. Algorithm~\ref{alg:hsrq3in} lists the inverse iteration solver. Following LAPACK, only a single iteration of \eqref{eq:inviteration} is computed for a given starting vector. After this single iteration, converged eigenvectors are separated from non-converged eigenvectors. New starting vectors are tried for the non-converged eigenvectors.

\begin{algorithm}[t]
\SetKwProg{Fn}{function}{}{end}
\caption{Robust Tiled Inverse Iteration for Real Eigenvalues}
  \label{alg:hsrq3in}
  \KwData{Hessenberg matrix $\boldsymbol{H} \in \mathbb{R}^{n \times n}$, eigenvalues $\boldsymbol{\Lambda} = [\lambda_1, \hdots, \lambda_m]^T \in \mathbb{R}^m$}
  \KwResult{$\boldsymbol{X} \in \mathbb{R}^{n \times m}$ such that $\boldsymbol{H} \boldsymbol{X} = \boldsymbol{X} \operatorname{diag}(\lambda_1, \hdots, \lambda_m)$}

  \Fn{\upshape \textsc{DHSRQ3IN}($\boldsymbol{H}, \boldsymbol{\Lambda}$)}{
    Allocate $\boldsymbol{X} \in \mathbb{R}^{n \times m}$\;
    Choose the starting vector $\boldsymbol{x}^{(0)} \gets 1/(\epsilon \|\boldsymbol{H} \|_\infty) \, [1, \hdots, 1]^T$\;
    Initialize $\boldsymbol{X}^{(0)}$ with $m$ repeated copies of $\boldsymbol{x}^{(0)}$\;
    \While{not converged}{
        Compute robustly $\boldsymbol{X}^{(1)} \gets \textsc{DHSRQ3}(\boldsymbol{H}, \boldsymbol{\Lambda}, \boldsymbol{X}^{(0)})$\;
        Split $\boldsymbol{X}^{(1)}$ into converged eigenvectors $\boldsymbol{X}_{\rm{c}}$ and non-converged eigenvectors $\boldsymbol{X}_{\rm{d}} \in \mathbb{R}^{n \times k}$\;
        Append $\boldsymbol{X}_{\rm{c}}$ to $\boldsymbol{X}$\;
        \If{$k = 0$}{
            Go to line 13\;  
        }
        Choose a new starting vector $\boldsymbol{x}^{(0)}$ orthogonal to previous choices\;
        Initialize $\boldsymbol{X}^{(0)}$ with $k$ repeated copies of $\boldsymbol{x}^{(0)}$\;      

    } 
    Sort the columns of $\boldsymbol{X}$ such that the order matches $\boldsymbol{\Lambda}$\;
    \Return $\boldsymbol{X}$\;
  }
\end{algorithm}

\subsection{Complex Shifts}\label{sec:complex-shifts}

The RQ factorization $\boldsymbol{H} - \lambda \boldsymbol{I} = \boldsymbol{R}\boldsymbol{Q}$ has complex factors $\boldsymbol{R}$ and $\boldsymbol{Q}$ if $\lambda$ is complex. The backward substitution with $\boldsymbol{R}$ then relies fully on complex arithmetic. Due to costly multiplications of complex scalars with complex vectors, Henry preferred the UL factorization over the RQ factorization for complex shifts, see Section~\ref{sec:rq}. This section presents two techniques which allow extending the inverse iteration solver \textsc{DHSRQ3IN} to support complex shifts at a reasonable computational cost. The first technique chooses the complex Givens rotation such that the reduction phase avoids multiplications of complex scalars with complex vectors. The second technique lowers the cost of the backward substitution by exploiting that most entries of $\boldsymbol{H} - \lambda \boldsymbol{I}$ are real in spite of a complex shift.
%

The reduction phase requires complex Givens rotations to compute the unitary $\boldsymbol{Q}$ factor. This paper adopts the Givens rotations applied by Beattie et al.~\cite[p. 6]{beattie2012note}
\begin{equation}\label{eq:cgivens}
\boldsymbol{G}_{j,\ell}^H =
\left[\begin{array}{cccc}
\boldsymbol{I}_{j-2} & \boldsymbol{0}           & \boldsymbol{0}                  & \boldsymbol{0} \\
\boldsymbol{0}       & c(j,{\ell}) & -s(j,{\ell})       & \boldsymbol{0} \\
\boldsymbol{0}       & s(j,{\ell}) &  \bar{c}(j,{\ell}) & \boldsymbol{0} \\
\boldsymbol{0}       & \boldsymbol{0}           & \boldsymbol{0}                  & \boldsymbol{I}_{n-j}
\end{array}\right] \in \mathbb{C}^{n \times n}
\end{equation}
where $c \in \mathbb{C}$ and $s \in \mathbb{R}$. Thereby most of the reduction phase corresponds to mixed real-complex multiplications. In view of the analysis of Givens rotations in floating-point arithmetic by Bindel et al.~\cite{bindel2002computing}, our implementation of this Givens rotation is numerically robust and takes care of underflow and overflow. 

Next we discuss the changes to Algorithm~\ref{alg:tiled-reduction1} \textsc{ReduceDiag} and Algorithm~\ref{alg:tiled-reduction2} \textsc{ReduceOffdiag}. The columns of the $\boldsymbol{R}$ factor are complex and so are the cross-over columns. We store a complex vector $\boldsymbol{v} = \boldsymbol{v}^{\rm{re}} + i \boldsymbol{v}^{\rm{im}}$ as adjacent columns $\begin{bmatrix}
\boldsymbol{v}^{\rm{re}} & \boldsymbol{v}^{\rm{im}}
\end{bmatrix}$. Using the complex Givens rotation \eqref{eq:cgivens}, the complex version of line 5 in Algorithm~\ref{alg:tiled-reduction2} \textsc{ReduceOffdiag}
\begin{displaymath}
\underbrace{\boldsymbol{v}(1:k)}_{\mathbb{C}^k} \gets \underbrace{c(j,\ell)}_{\mathbb{C}} \underbrace{\boldsymbol{h}(1:k,j)}_{\mathbb{R}^k} + \underbrace{s(j,\ell)}_{\mathbb{R}} \underbrace{\boldsymbol{v}(1:k)}_{\mathbb{C}^k}
\end{displaymath}
can be realized
\begin{align*}
\boldsymbol{v}^{\rm{re}}(1:k) &\gets c^{\rm{re}}(j,\ell) \boldsymbol{h}(1:k,j) + s(j,\ell) \boldsymbol{v}^{\rm{re}}(1:k) \\
\boldsymbol{v}^{\rm{im}}(1:k) &\gets c^{\rm{im}}(j,\ell) \boldsymbol{h}(1:k,j) + s(j,\ell) \boldsymbol{v}^{\rm{im}}(1:k).
\end{align*}
The lines 6--7 in \textsc{ReduceDiag} can be realized in a similar fashion. The flop count for the complex versions of \textsc{ReduceDiag} and \textsc{ReduceOffdiag} doubles compared to their real counterparts and is $3n^2 + \mathcal{O}(n)$ and $6nk + \mathcal{O}(n)$, respectively.

Next we discuss the changes to the backward substitution phase. Systems with a complex shift yield a complex solution. Aiming for a robust backward substitution, every complex solution vector is associated with a single scaling factor. In other words, the real and the imaginary part are scaled alike. Then a robust \textsc{Solve} task addresses $\boldsymbol{R}_{\ell} \boldsymbol{y}_{\ell} = \gamma_{\ell} \boldsymbol{b}_{\ell}$ where $\gamma_{\ell} \in (0,1]$ and all other quantities are complex. If the (small) triangular system matrices $\boldsymbol{R}_{\ell}$ are computed explicitly, a conversion to a complex datatype allows the robust solution of this system with a call to \textsc{CLATRS} for every complex right-hand side. \textsc{CLATRS} is available in LAPACK 3.9.0. and is the complex counterpart of \textsc{DLATRS}, see Section~\ref{sec:overflow}. Thereby Algorithm~\ref{alg:robust-solve} naturally generalizes to complex arithmetic. The flop count is $15n^2 + \mathcal{O}(n)$ per right-hand side.

\begin{table*}
  \caption{Flops approximations for a single shift/eigenvector segment of length $n$ and assuming square tiles. Flops due to overflow protection (norm computations, overflow protection logic, numerical scaling) are disregarded.}
  \label{tab:flops-vector}
  \centering
  \begin{tabular}{lcc}
    \toprule
    Routine                &  Real eigenvector & Complex eigenvector \\
    \midrule
    \textsc{ReduceDiag}    & $1.5 n^2 + \mathcal{O}(n)$      & $3n^2 + \mathcal{O}(n)  $ \\
    \textsc{ReduceOffdiag} & $3 n^2 + \mathcal{O}(n)$        & $6n^2 + \mathcal{O}(n)  $ \\
    \textsc{Solve}         & $3.5 n^2  + \mathcal{O}(n)$     & $15n^2 + \mathcal{O}(n)$   \\
    \textsc{Update}        & $2n^2 + \mathcal{O}(n+n)$       & $4n^2 + \mathcal{O}(n)$ \\
    \textsc{Backtransform} & $6(n-1)$                        & $20(n-1)$   \\
    \bottomrule
  \end{tabular}
\end{table*}

An analysis of the complex counterpart of the linear update \eqref{eq:robust-update}
\begin{align*}
\delta_{\ell}^{-1} \boldsymbol{b}_{\ell}(1:j-1) \gets \beta_{\ell}^{-1} \boldsymbol{b}_{\ell}(1:j-1) - \hspace{6cm}&\\
\begin{bmatrix} 
\boldsymbol{h}(1:j-1,j-1) - \lambda_{\ell} \boldsymbol{e}_{j-1} & \boldsymbol{H}(1:j-1,j:k-2) & \tilde{\boldsymbol{r}}_{\ell}(1:j-1)\\
\end{bmatrix}  \qquad&\\
\left( \boldsymbol{G}_2^T
\begin{bmatrix}
0 \\
\alpha_{\ell}^{-1} \boldsymbol{x}_{\ell}(j:k-1)
\end{bmatrix}
\right)&
\end{align*}

\noindent reveals the potential for mixed real-complex arithmetic. Analogously to real linear update, the block structure of the system matrix suggests three block operations. 
The first block $\boldsymbol{h}(1:j-1,j-1) - \lambda_{\ell} \boldsymbol{e}_{j-1}$ and the third block $\tilde{\boldsymbol{r}}_{\ell}(1:j-1)$ issue multiplications of a complex scalar with a complex vector and can be emulated using real arithmetic. The second block $\boldsymbol{H}(1:j-1,j:k-2)$ is real and requires the multiplication of a real matrix with a complex vector. If complex quantities are stored in interleaved storage, that is, the real and imaginary parts of a complex vector are stored in adjacent columns, the multiplication of a real matrix $\boldsymbol{H}$ and a complex vector $\boldsymbol{z} = \boldsymbol{u} + i\boldsymbol{v}$ can be realized as $\boldsymbol{H} \begin{bmatrix} \boldsymbol{u} & \boldsymbol{v}\end{bmatrix}$. Hence, the computationally expensive part of the linear update can be realized with a wide \textsc{DGEMM} operation if many right-hand sides are computed simultaneously. The flop count of a linear update with an $m$-by-$n$ matrix $\boldsymbol{H}$ is approximately $4mn + \mathcal{O}(m + n)$ per complex right-hand side.

The complex solution $\boldsymbol{x}_{\ell}$ is backtransformed analogously to \eqref{eq:backtransform} using complex arithemtic. The flop count sums to $20n+\mathcal{O}(1)$. An overview of all flop counts is listed in Table~\ref{tab:flops-vector}. The computational cost per column is approximately the same for a real eigenvector and a complex eigenvector comprising two columns.

Coupling the complex routines of all task types results in \textsc{CHSRQ3}, the complex counterpart of \textsc{DHSRQ3}. Analogously to LAPACK, the starting vector and the convergence criterion are chosen identically for real and complex eigenvalues. \textsc{CHSRQ3} allows generalizing the inverse iteration routine \textsc{DHSRQ3IN} to handle \emph{any} selection of eigenvalues, see Algorithm~\ref{alg:hsrq3infinal}. The selection of eigenvalues is split into real eigenvalues and complex eigenvalues. This requires additional tracking of the affiliation between computed eigenvectors and selected eigenvalues. Recall that complex eigenvectors occur in complex conjugate pairs. Hence, for each selected complex conjugate pair of eigenvalues, only one eigenvector has to be computed. The other one can be obtained for free by complex conjugation. \textsc{HSRQ3IN} assumes that the provided eigenvalues exploit this.

The real/complex eigenvectors are computed by successive calls to the real/complex version of \textsc{HSRQ3IN}.  The successive computation of real/complex eigenvectors is justified because the storage requirement quickly limits the problem sizes solvable with \textsc{HSRQ3IN}. An analysis of the storage requirement is given in the next section. 

\begin{algorithm}[t]
\SetKwProg{Fn}{function}{}{end}
\caption{Robust Tiled Inverse Iteration}
  \label{alg:hsrq3infinal}
  \KwData{Hessenberg matrix $\boldsymbol{H} \in \mathbb{R}^{n \times n}$, tile size $b_{\rm r}$ with $n = Nb_{\rm r}$, eigenvalues $\boldsymbol{\Lambda} = [\lambda_1, \hdots, \lambda_m]^T \in \mathbb{C}^m$ such that only one complex eigenvalue of a complex conjugate pair is included in $\boldsymbol{\Lambda}$}
  \KwResult{$\boldsymbol{X} \in \mathbb{C}^{n \times m}$ such that $\boldsymbol{H} \boldsymbol{X} = \boldsymbol{X} \operatorname{diag}(\lambda_1, \hdots, \lambda_m)$}

  \Fn{\upshape \textsc{HSRQ3IN}($\boldsymbol{H}, \boldsymbol{\Lambda}$)}{
    Sort $\boldsymbol{\Lambda}$ into $\boldsymbol{\Lambda}_{\rm{sorted}} = \begin{bmatrix}\boldsymbol{\Lambda}_{\rm{real}} & \boldsymbol{\Lambda}_{\rm{complex}}\end{bmatrix}$ so that the first $m_1$ eigenvalues are real\;
    Define the maximum workspace size for the cross-over columns as $w_{\max}$\;
    $g \gets \lfloor w_{\max} / (2nN) \rfloor$\tcp*{Group size}
    Allocate workspaces for the Givens rotations and the cross-over columns\;
    \For{$\ell \gets 1:2g:m_1$}{
      \tcp{Real eigenvectors}
      $\mathcal{L} \gets \ell : \max\{\ell + 2g, m_1\}$\;
      Task-parallel $\boldsymbol{X}(\mathcal{L}) \gets \textsc{DHSRQ3IN}(\boldsymbol{H},\boldsymbol{\Lambda}_{\rm{sorted}}(\mathcal{L}))$\;
    }
    \For{$\ell \gets m_1 + 1:g:m$}{
      \tcp{Complex eigenvectors}
      $\mathcal{L} \gets \ell : \max\{\ell + g, m_2\}$\;
      Task-parallel $\boldsymbol{X}(\mathcal{L}) \gets \textsc{CHSRQ3IN}(\boldsymbol{H},\boldsymbol{\Lambda}_{\rm{sorted}}(\mathcal{L}))$\;
    }

    Sort the columns of $\boldsymbol{X}$ so that the order matches $\boldsymbol{\Lambda}$\;
    \Return $\boldsymbol{X}$\;
  }
\end{algorithm}

\subsection{Storage Requirement Analysis}
Algorithm~\ref{alg:hsrq3in} requires the recording of the Givens rotations and the cross-over columns computed in the reduction phase. If the Hessenberg matrix is $n$-by-$n$, $n$ Givens rotations including the first padded entry have to be recorded for each computed eigenvector. For a real eigenvector, the storage requirement of all Givens rotations is $2n$ quantities if both $s$ and $c$ of the Givens rotations \eqref{eq:givens} are recorded. This storage requirement can be reduced to $n$ quantities by storing the Givens rotations compactly as proposed by Stewart~\cite{stewart1976economical}. Stewart exploits the relationship $c^2 + s^2 = 1$ and stores either $c$ or $s$, depending on which quantity is normwise closer to 1. Then other quantity can be computed in a numerically safe way. For a complex eigenvector, the storage of all complex Givens rotations \eqref{eq:cgivens} including the first padded entry comprises $3n$ quantities. These quantities cover the real and imaginary part of $c$ and the real $s$. Stewart's idea can be applied twice, reducing the storage requirement to $2n$ per complex eigenvector.

The storage requirement of the cross-over columns depends on the tiling of $\boldsymbol{H}$. The partitioning of $\boldsymbol{H}$ into an $N$-by-$N$ grid with tiles of size $b_{\rm r} \times b_{\rm r}$ requires the storage of $N$ (potentially complex) cross-over columns. The storage requirement of the cross-over columns sums to $\sum_{i = 1}^{N}(ib_{\rm{r}}) = \mathcal{O}(nN)$ for every eigenvector. If $m$ columns (1 column per real or 2 columns per complex eigenvector) are computed, the storage of the cross-over columns is $\mathcal{O}(nNm)$ and can quickly exceed the memory available on a compute node. This problem can be addressed by computing the eigenvectors in groups. Algorithm~\ref{alg:hsrq3infinal} realizes this. Only workspace necessary for storing the cross-over columns and the Givens rotations of a single group has to be allocated. Once a group has been computed, the workspace can be reused for the next group. Since complex eigenvectors and cross-over column are stored in interleaved storage, the storage requirement is doubled compared to the real computation. Line 4 calculates the group size for the complex case requiring two columns per cross-over column. To fully harness the available workspace, the real computation doubles the group size (lines 6--8).

\section{Numerical experiments}\label{sec:experiments}

This section describes how the numerical experiments were set up and executed and presents the results.

\subsection{Execution Environment}

\paragraph{Hardware}
The experiments are run on an Intel Xeon Gold 6132 (Skylake) node where dynamic frequency scaling is enabled. This node has 2 NUMA islands with 14 cores each. In double-precision arithmetic the theoretical peak performance is 83.2 GFLOPS/s per core and 2329.6 GFLOPS/s per node. The available memory is 192 GB RAM. The memory bandwidth was measured at 12.7 GB/s for one core and 162 GB/s for a full node using the STREAM triad benchmark.

\paragraph{Software and configuration}
The software is built with the Intel compiler 19.0.1.144 where the optimization level is set to \texttt{-O2}, AVX-512 instructions are enabled and interprocedural optimizations \texttt{-ipo} are activated. We link against the MKL 2019.1.144 BLAS implementation. OpenMP threads are bound to physical processing units by setting \texttt{KMP\_AFFINITY} to \texttt{compact}.

In the following we describe the routines and their configuration used in the numerical experiments. The first routine targets shifted Hessenberg systems and solves $(\boldsymbol{H} - \lambda_{\ell} \boldsymbol{I})\boldsymbol{x}_{\ell} = \alpha_{\ell}\boldsymbol{e}$. The matrix $\boldsymbol{H}$ is real and upper Hessenberg, the shift $\lambda_{\ell}$ is real or complex, $\boldsymbol{e}$ is the vector with all ones and $\alpha_{\ell} \in (0,1]$ is a scaling factor. The next four routines target the computation of eigenvectors by inverse iteration. The routines are supplied with $\| \boldsymbol{H}\|_\infty \epsilon \boldsymbol{e}$ as the starting vector, see Section~\ref{sec:starting-vector}. This starting vector leads to convergence in one iteration in all of our numerical experiments.

\begin{itemize}

\item \textsc{\{C,D\}HSRQ3}. This robust routine was introduced in this work and solves $(\boldsymbol{H} - \lambda_{\ell} \boldsymbol{I}) \boldsymbol{x}_{\ell} = \alpha_{\ell} \boldsymbol{e}$. It generalizes the RQ approach originally proposed by Henry. The reduction phase records the Givens rotations necessary to compute the RQ factorizations for all shifts. The backward substitution phase utilizes level-3 BLAS for the linear updates.

\item \textsc{RQIN} (Henry). This routine extends the shifted Hessenberg system solver by Henry~\cite[Algorithm~2]{henry1994techreport} to the computation of eigenvectors. It solves $(\boldsymbol{H} - \lambda_{\ell} \boldsymbol{I})\boldsymbol{x}_{\ell} = \| \boldsymbol{H}\|_\infty \epsilon \boldsymbol{e}$ through an RQ factorization for every eigenvalue and normalizes every eigenvector after the backtransform. The core of this routine corresponds to Algorithm~\ref{alg:henry}. While Henry uses the RQ decomposition only for real shifts, the numerical experiments use the RQ approach both for real and complex shifts. The matrix $\boldsymbol{H}$ is not overwritten.

\item \textsc{ULIN} (Henry). This routines solves $(\boldsymbol{H} - \lambda_{\ell} \boldsymbol{I})\boldsymbol{x}_{\ell} = \| \boldsymbol{H}\|_\infty \epsilon \boldsymbol{e}$ through an UL factorization for every shift and, in a final step, normalizes the computed eigenvectors. Henry~\cite[Section~4]{henry1994techreport} introduced the UL approach for complex shifts to avoid costly complex-complex multiplications. Since the UL approach overwrites $\boldsymbol{H}$, the original matrix $\boldsymbol{H}$ is copied when more than one system is solved. 

\item \textsc{DHSEIN}. LAPACK 3.9.0 contains the driver routine \textsc{DHSEIN} for successively computing selected left and/or right eigenvectors of a real upper Hessenberg matrix. \textsc{DHSEIN} calls \textsc{DLAEIN} for computing a single eigenvector by inverse iteration. The routine is supplied with $\epsilon\boldsymbol{e}$ as the user-defined starting vector (\textsc{INITV='U'}). For each shift, the matrix $\boldsymbol{B} = \boldsymbol{H}-\lambda \boldsymbol{I}$ is explicitly constructed in a workspace. Then $\boldsymbol{B} \boldsymbol{x} = \alpha (\| \boldsymbol{H}\|_\infty \epsilon \boldsymbol{e})$ is solved through an LU factorization with partial pivoting. In all numerical experiments conducted here, this initial guess lead to convergence in the first iteration. In other words, \textsc{DHSEIN} effectively solves a single shifted Hessenberg system through an LU decomposition with partial pivoting for each eigenvector. 

\item \textsc{HSRQ3IN}. This driver routine, listed in Algorithm~\ref{alg:hsrq3infinal}, splits real and complex eigenvalues and computes the corresponding eigenvectors by successive calls to \textsc{DHSRQ3IN} and \textsc{CSRQ3IN} . It solves $(\boldsymbol{H} - \lambda_{\ell} \boldsymbol{I})\boldsymbol{x}_{\ell} = \alpha_{\ell} (\| \boldsymbol{H}\|_\infty \epsilon \boldsymbol{e})$ and normalizes the eigenvectors with respect to the Euclidean norm before the backtransform.
\end{itemize}

\subsection{Test Problems}
The numerical experiments use two test problems. The \textbf{first test problem} is designed to have known, well-separated eigenvalues and computes the corresponding eigenvectors. This experiment controls the ratio of real/complex eigenvalues and allows us to examine the cost of complex arithmetic. For this purpose, a quasi-triangular matrix $\boldsymbol{T} \in \mathbb{R}^{n \times n}$ is constructed where the eigenvalues are placed as 1-by-1 or 2-by-2 blocks on the diagonal of $\boldsymbol{T}$. If the $k$-th eigenvalue is real, then the 1-by-1 diagonal block is $t(k,k) = k$. Complex eigenvalues occur in complex conjugate pairs and correspond to 2-by-2 blocks. Such a 2-by-2 block is set to
\begin{equation}
\boldsymbol{T}(k:k+1,k:k+1) =
\begin{bmatrix}
k & k \\
-k & k \\
\end{bmatrix}
\in \mathbb{R}^{2 \times 2}
\end{equation}
and corresponds to the eigenvalues $k + ik$ and $k-ik$. This choice of diagonal blocks ensures that all eigenvalues are well-separated. In particular, the case with 100\% real eigenvalues yields an upper triangular matrix with eigenvalues $1, 2,\hdots, n$. Assuming that $n$ is even, the case with 100\% complex eigenvalues yields a matrix with only 2-by-2 blocks on the diagonal and eigenvalues $1 \pm i, 3 (1 \pm i), \hdots, (n-1)(1\pm i)$. The remaining superdiagonal entries are random in $(0,1]$.

The matrix $\boldsymbol{T}$ is then transformed into a Hessenberg matrix through an orthogonal similarity transformation. For this purpose, two orthogonal transformations are applied. First, a random Householder matrix is constructed, $\boldsymbol{Q}_0 = \boldsymbol{I} - 2\boldsymbol{v} \boldsymbol{v}^T$, where $\boldsymbol{v}$ is a random unit norm vector. By applying the (symmetric) Householder matrix, a dense matrix $\boldsymbol{A} = \boldsymbol{Q}_0 \boldsymbol{T} \boldsymbol{Q}_0$ is computed. Second, $\boldsymbol{A}$ is reduced to Hessenberg form through the LAPACK routine DGEHRD $\boldsymbol{H}_1 \gets \boldsymbol{Q}_1^T \boldsymbol{A} \boldsymbol{Q}_1$. Together, $\boldsymbol{H}_1$ is given by $\boldsymbol{H}_1 \gets (\boldsymbol{Q}_0 \boldsymbol{Q}_1)^T \boldsymbol{T} (\boldsymbol{Q}_1 \boldsymbol{Q}_0)$. The numerical routines receive the exact eigenvalues as input parameter.

The \textbf{second test problem} solves shifted Hessenberg systems and aims at quantifying the overhead from overflow protection. Two systems are constructed: the ``bad'' system requires frequent numerical scaling system, whereas the ``good'' system never requires numerical scaling.

The bad system constructs the Hessenberg matrix $\boldsymbol{H}_2 = \boldsymbol{R}_2 \boldsymbol{Q}_2 + \gamma \boldsymbol{I}$ where $\boldsymbol{R}_2 \in \mathbb{R}^{n \times n}$ and $\boldsymbol{Q}_2 \in \mathbb{R}^{n \times n}$ are given by
\begin{displaymath}
r_2(i,j) = 
\begin{cases}
n-i+1 & i = j \\
-n & j > i \\
0   & i > j
\end{cases},
\end{displaymath}
\begin{displaymath}
q_2(i,j) = 
\begin{cases}
-1 & (i-1 = j) \text{ or } (i = 1 \text{ and }j = n)\\
0   & \text{otherwise}.
\end{cases}
\end{displaymath}
As an example, consider how $\boldsymbol{H}_2 - \gamma \boldsymbol{I}$ is constructed for $n = 6$ by
\begin{align*}
\boldsymbol{H}_2 - \gamma \boldsymbol{I}&=
\begin{bmatrix}
6  & 6  & 6 & 6 & 6 & -6 \\
-5 & 6  & 6 & 6 & 6 & 0 \\
   & -4 & 6 & 6 & 6 & 0 \\
   &   & -3 & 6 & 6 & 0 \\
   &   &   & -2 & 6 & 0 \\
   &   &   &   & -1 & 0
\end{bmatrix}\\
&=
\boldsymbol{R}_2 \boldsymbol{Q}_2 \\
&=
\begin{bmatrix}
6  & -6  & -6 & -6 & -6 & -6 \\
 & 5  & -6 & -6 & -6 & -6 \\
   &  & 4 & -6 & -6 & -6 \\
   &   &  & 3 & -6 & -6 \\
   &   &   &  & 2 & -6 \\
   &   &   &   &   & 1
\end{bmatrix}
\begin{bmatrix}
0  & 0  & 0 & 0 & 0 & -1 \\
-1 & 0  & 0 & 0 & 0 & 0 \\
   & -1 & 0 & 0 & 0 & 0 \\
   &   & -1 & 0 & 0 & 0 \\
   &   &   & -1 & 0 & 0 \\
   &   &   &   & -1  & 0
\end{bmatrix}.
\end{align*}

The numerical experiments solve the same shifted system $(\boldsymbol{H}_2 - \gamma \boldsymbol{I}) \boldsymbol{x}_{\ell} = \boldsymbol{e}$ where $\gamma = 2$ repeatedly without exploiting that the shift is shared. Kjelgaard Mikkelsen~\cite{mikkelsen2020well} has shown that the matrix $\boldsymbol{R}_2$ introduces quick growth to the solution vectors during the backward substitution.

The good system constructs the Hessenberg matrix $\boldsymbol{H}_3 = \boldsymbol{R}_3 \boldsymbol{Q}_2 + \gamma \boldsymbol{I}$ where $\boldsymbol{R}_3 \in \mathbb{R}^{n \times n}$ is given by
\begin{displaymath}
r_3(i,j) = 
\begin{cases}
n + 1 - i & i = j \\
\frac{1}{2} & j > i \\
0   & i > j.
\end{cases}
\end{displaymath}
The numerical experiments solve $(\boldsymbol{H}_3 - \gamma \boldsymbol{I})\boldsymbol{x}_{\ell} = \boldsymbol{e}$ without harnessing that the shift is shared. Following Kjelgaard Mikkelsen~\cite{mikkelsen2020well}, the solution of $\boldsymbol{R}_3 \boldsymbol{y} = \boldsymbol{e}$ with backward substitution never requires numerical scaling to avoid overflow.

\subsection{Results}\label{sec:results}
This section presents the results of four numerical experiments. The first experiment concerns the sequential runtimes of the inverse iteration solvers and compares the existing approaches \textsc{DHSEIN} (LAPACK), \textsc{RQIN}/\textsc{ULIN} (Henry) with \textsc{HSRQ3IN} introduced in this paper. The second experiment aims at identifying bottlenecks in the implementation and analyzes what fraction each computational phase contributes to the total runtime. The third experiment addresses the parallel scalability of \textsc{HSRQ3IN}. The fourth experiment quantifies the cost of robustness.

\paragraph{Serial Comparison}
\begin{figure}[t]
  \centering
  \begin{minipage}{\textwidth}
  \centering
  \includegraphics[width=.7\textwidth]{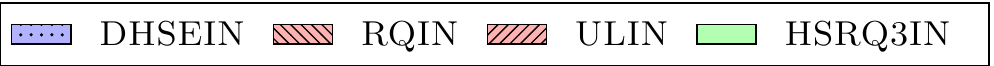}  
  \end{minipage}
  \begin{minipage}[b]{.625\textwidth}
  \centering
  \includegraphics[width=\textwidth]{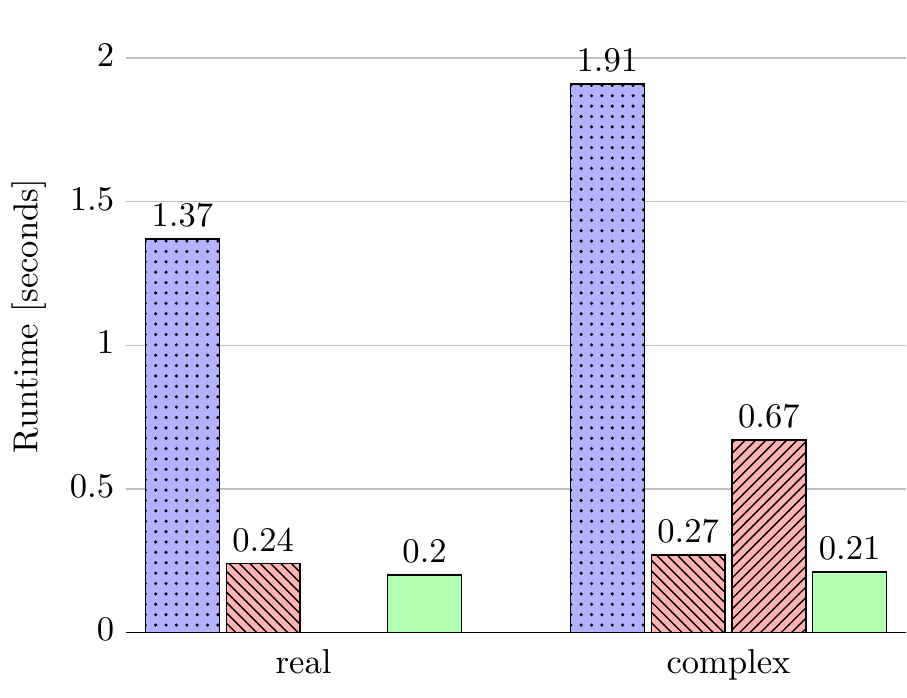}  
  \end{minipage}
  \begin{minipage}[b]{.625\textwidth}
  \centering
  \includegraphics[width=\textwidth]{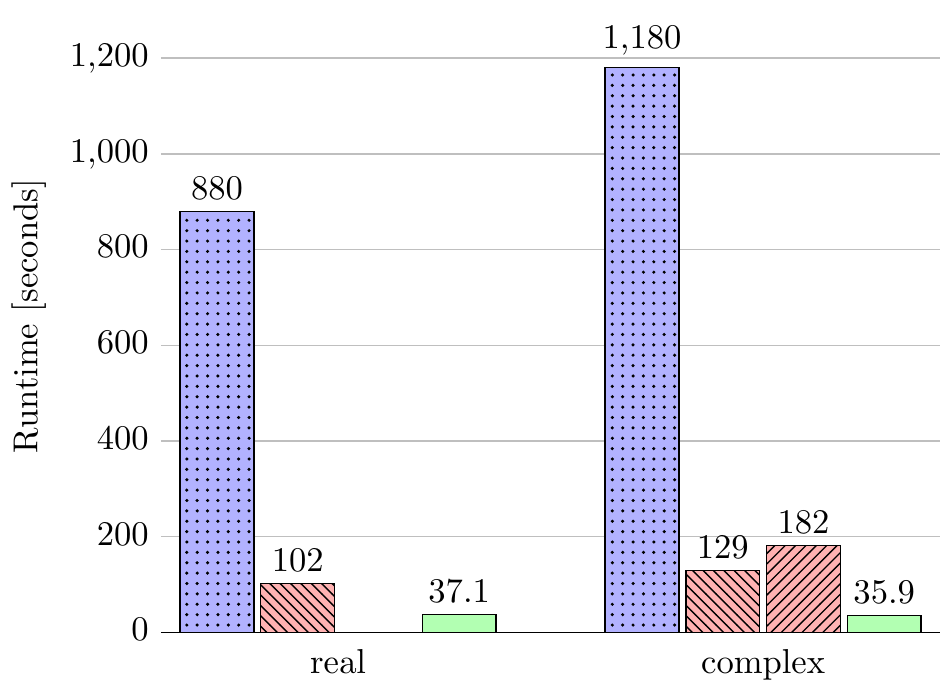}  
  \end{minipage}
  \caption{Sequential runtimes on $\boldsymbol{H}_1$ computing two columns (top) and 1500 columns (bottom) for $n = 10000$.}
  \label{fig:sequential}
\end{figure}

The first experiment compares the sequential runtimes of the inverse iteration routines \textsc{DHSEIN}, \textsc{RQIN}, \textsc{ULIN} and \textsc{HSRQ3IN} on $\boldsymbol{H}_1$ where $n = 10000$. Numerical scaling is never triggered and a single iteration suffices to satisfy the convergence criterion for every eigenvector. Figure~\ref{fig:sequential} (top) shows the runtimes for computing two columns, either two real vectors or a single complex vector (storing the real and the imaginary part in two columns). In the latter case, the UL approach overwrites the input matrix and, hence, does not require any data copies. For all solvers the computation of the complex vector is at least as expensive as the computation of two real vectors.

Figure~\ref{fig:sequential} (bottom) compares the runtimes for computation of 1500 columns, either 1500 real vectors or 750 complex vectors. The timing of the UL approach includes the overhead of 749 copies of the input matrix. The runtime gap between \textsc{DHSEIN} and the RQ factorization-based solvers widens when the number of right-hand sides is increased. Increasing the number of right-hand sides allows reusing data. Since the RQ decomposition-based solvers do not overwrite the input matrix, the computation may benefit from temporal locality. \textsc{HSRQ3IN} outperforms \textsc{RQIN}, which can be attributed to the matrix--matrix multiplications (level-3 BLAS) in the backward substitution phase. Complex eigenvectors are more expensive than real eigenvectors for \textsc{DHSEIN} and \textsc{RQIN}. \textsc{HSRQ3IN}, by contrast, performs similarly. The next experiment aims at investigating the underlying cause.

\paragraph{Analysis}
This experiment decomposes the computational cost of \textsc{HSRQ3IN} and \textsc{RQIN} and thereby analyzes the ratio of the three computational phases (reduction, backward substitution, backtransform). The experiment setup is identical to the one in Figure~\ref{fig:sequential} (right) and uses $\boldsymbol{H}_1$ with $n=10000$ for the computation of either 1500 real or 750 complex eigenvectors.

\begin{figure}
  \begin{minipage}[b]{.375\textwidth}
    \includegraphics[width=\textwidth]{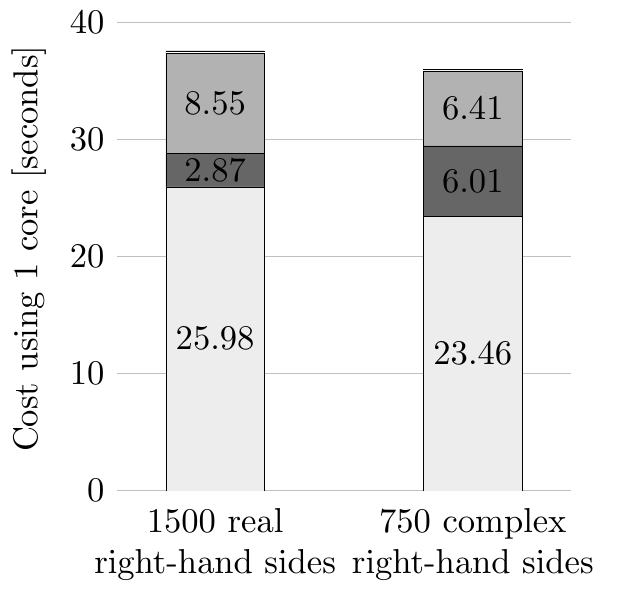}
  \end{minipage}
  \begin{minipage}[b]{.6\textwidth}
    \includegraphics[width=\textwidth]{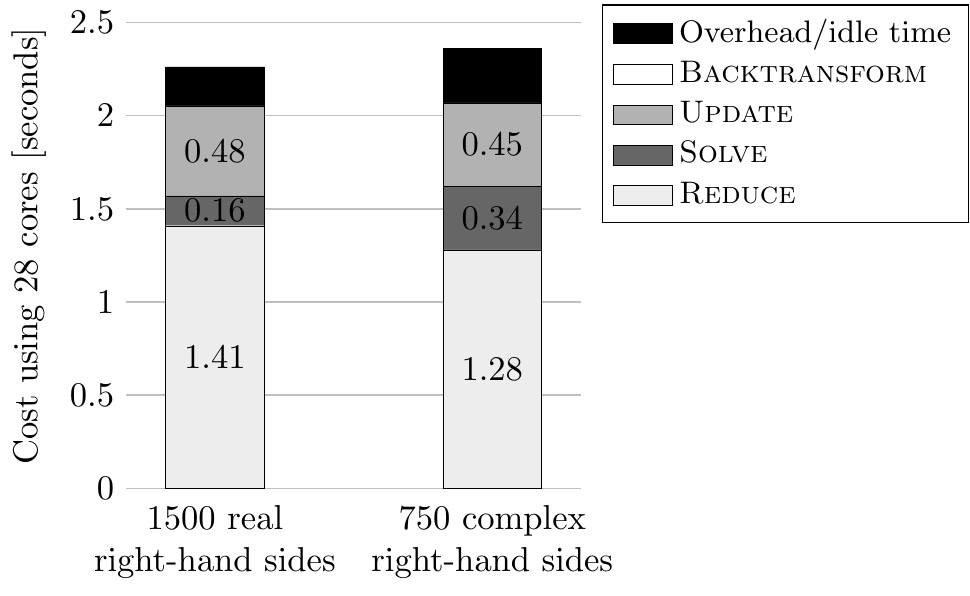}
  \end{minipage}
  \caption{Runtime decomposition of a sequential (left) and parallel (right) run on $\boldsymbol{H}_1$ for $n = 10000$.}
  \label{fig:analysis}
\end{figure}

A run of \textsc{RQIN} on 1500 real eigenvectors spends 46\% of the runtime in the reduction phase (lines 3, 8--9 in Algorithm~\ref{alg:henry}) and 53\% of the runtime in the backward substitution phase (lines 4--7 in Algorithm~\ref{alg:henry}). This ratio is approximately in line with the flop distribution of these two phases. When \textsc{RQIN} computes 750 complex eigenvectors, the reduction phase constitutes 36\% and the backward substitution phase 63\% of the runtime. Thus, the majority of the time is spent on backward substitution.

The runtime decompositions of \textsc{HSRQ3IN} are shown in Figure~\ref{fig:analysis}. The runtime is split into the contribution of each task type to the total compute time for a sequential (left) and a parallel (right) run.
Between 53\% and 68\% of the runtime is spent on \textsc{Reduce} tasks. The runtime difference between the real and the complex runs is due to the different amount of Givens rotations computed during the run. Since a series of Givens rotation is computed per eigenvector, the complex run computes only half the number of Givens rotations, but requires mixed real-complex multiplications. The backward substitution phase contributes with \textsc{Solve} and \textsc{Update} tasks. The complex runs spend approximately double the time on \textsc{Solve} tasks than the real runs. This can be attributed to the complex-complex multiplications during the small backward substitutions. The runtime difference of \textsc{Update} tasks is due to the application of the Givens rotations. Analogously to the \textsc{Reduce} tasks, the complex runs apply only half the number of Givens compared to the real runs. The backtransform phase makes a negligible contribution to the total runtime. The runtime decompositions of \textsc{RQIN} and \textsc{HSRQ3IN} suggest that the reduction phase has become the new bottleneck of the revised RQ approach.

The parallel speedup ranges in 14--18 for all task types. Due to dynamic frequency scaling, the best possible parallel speedup on the test node is 18.4. Idle cores during the parallel runs contribute to the overhead/idle time when there are not enough tasks available for being scheduled.

\begin{figure}
  \begin{minipage}[b]{.475\textwidth}
    \includegraphics[width=\textwidth]{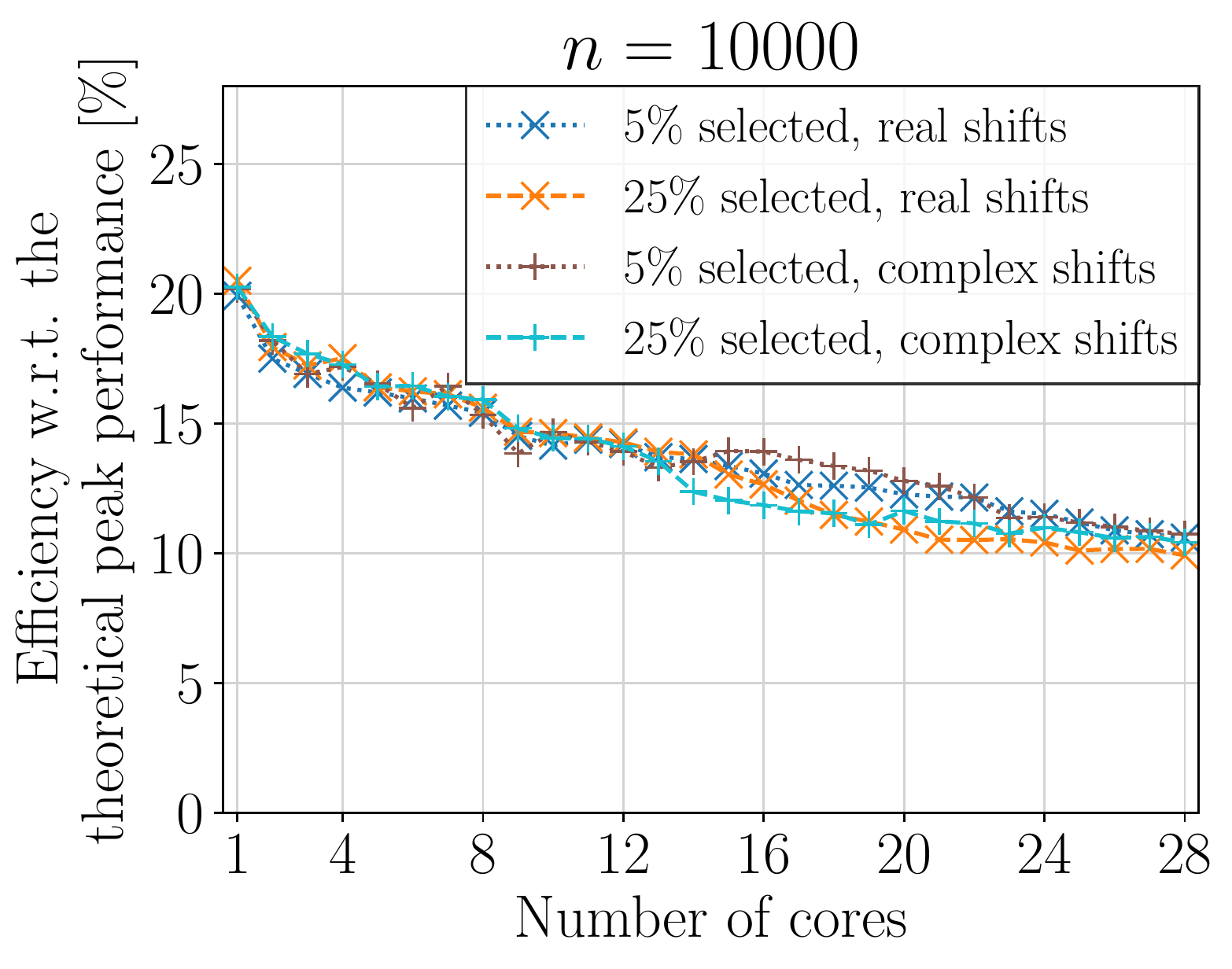}
  \end{minipage}
  ~
  \begin{minipage}[b]{.475\textwidth}
    \includegraphics[width=\textwidth]{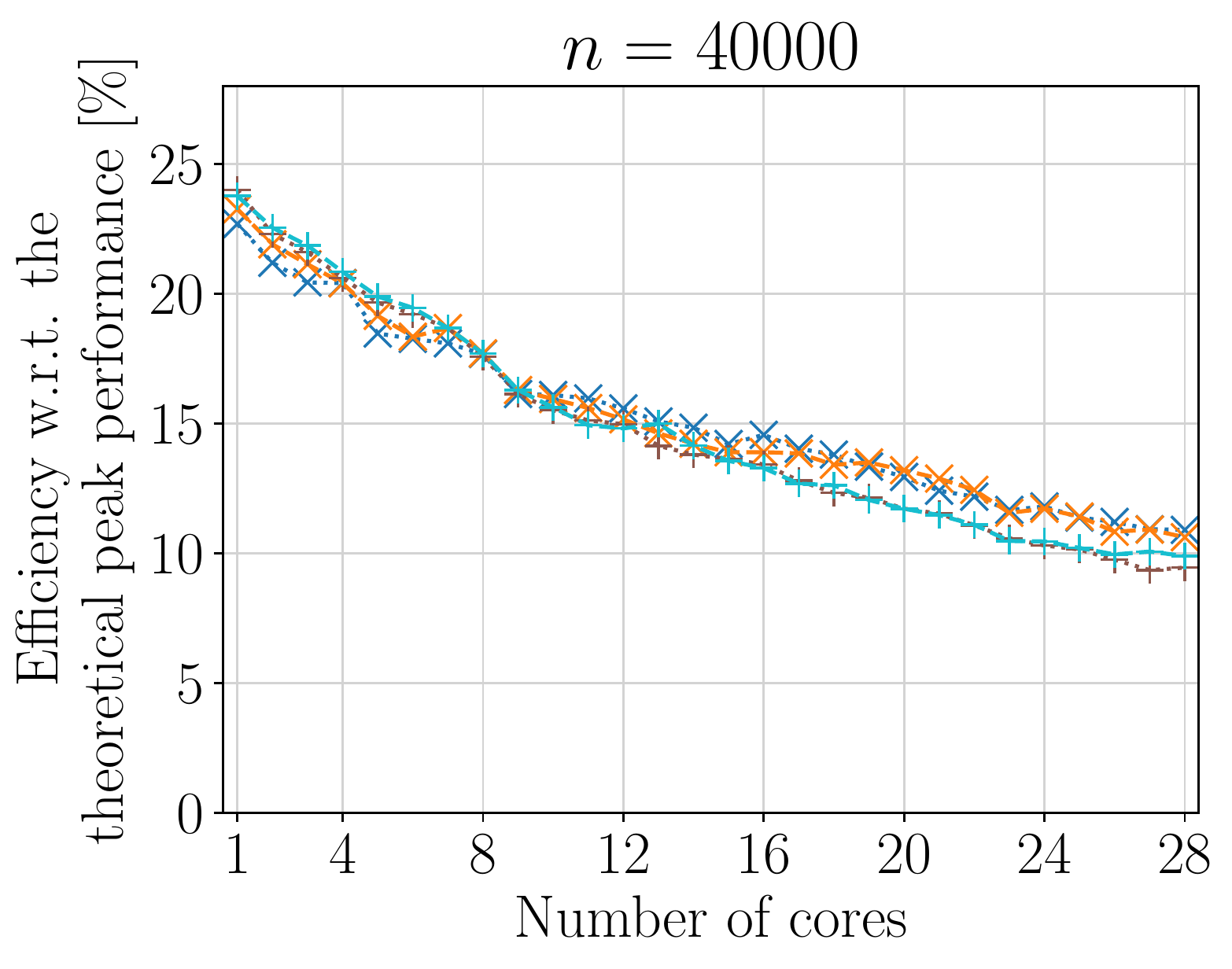}
  \end{minipage}
  \begin{minipage}[b]{.475\textwidth}
    \includegraphics[width=\textwidth]{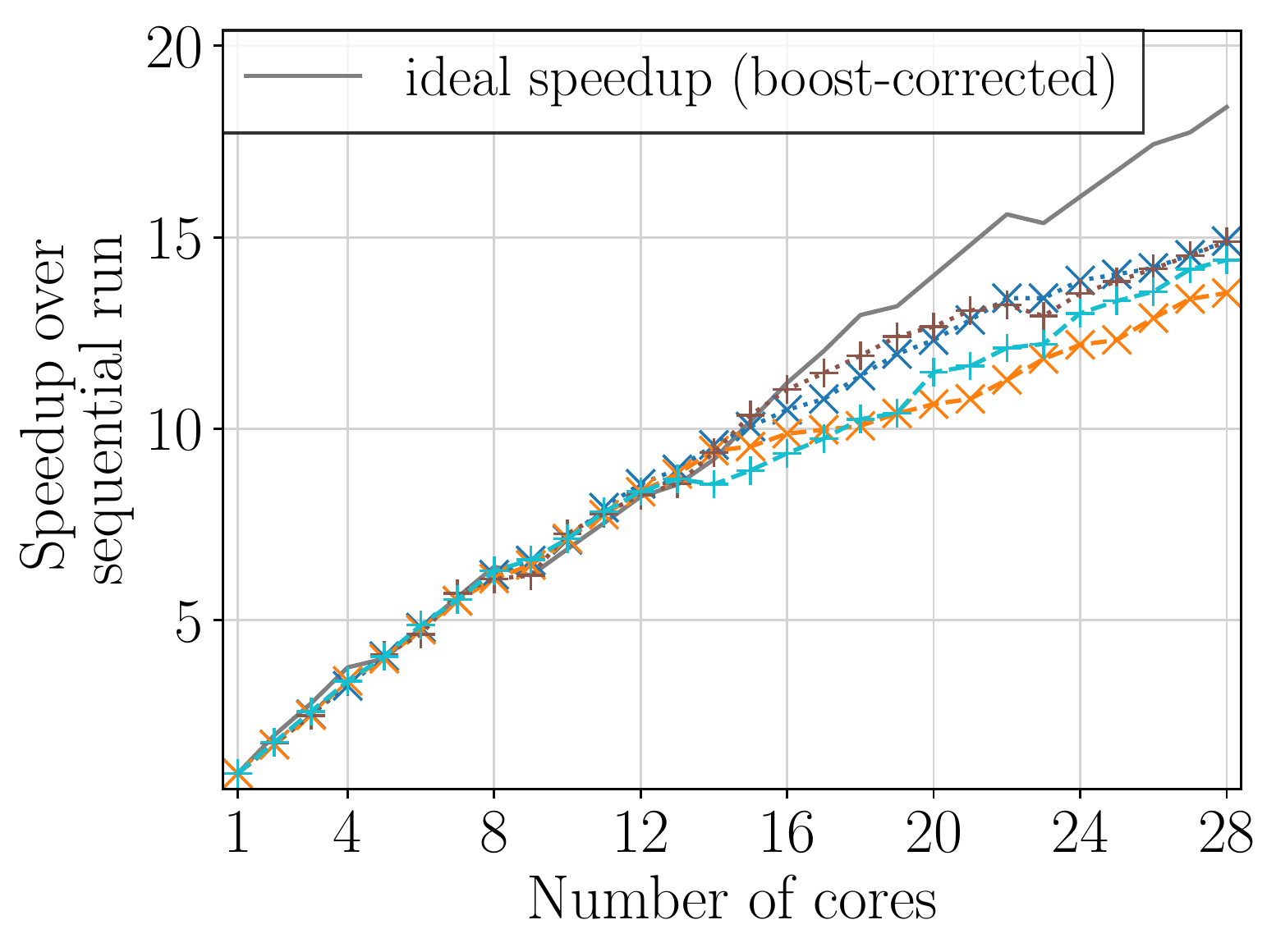}
  \end{minipage}
  ~\hfill
  \begin{minipage}[b]{.475\textwidth}
    \includegraphics[width=\textwidth]{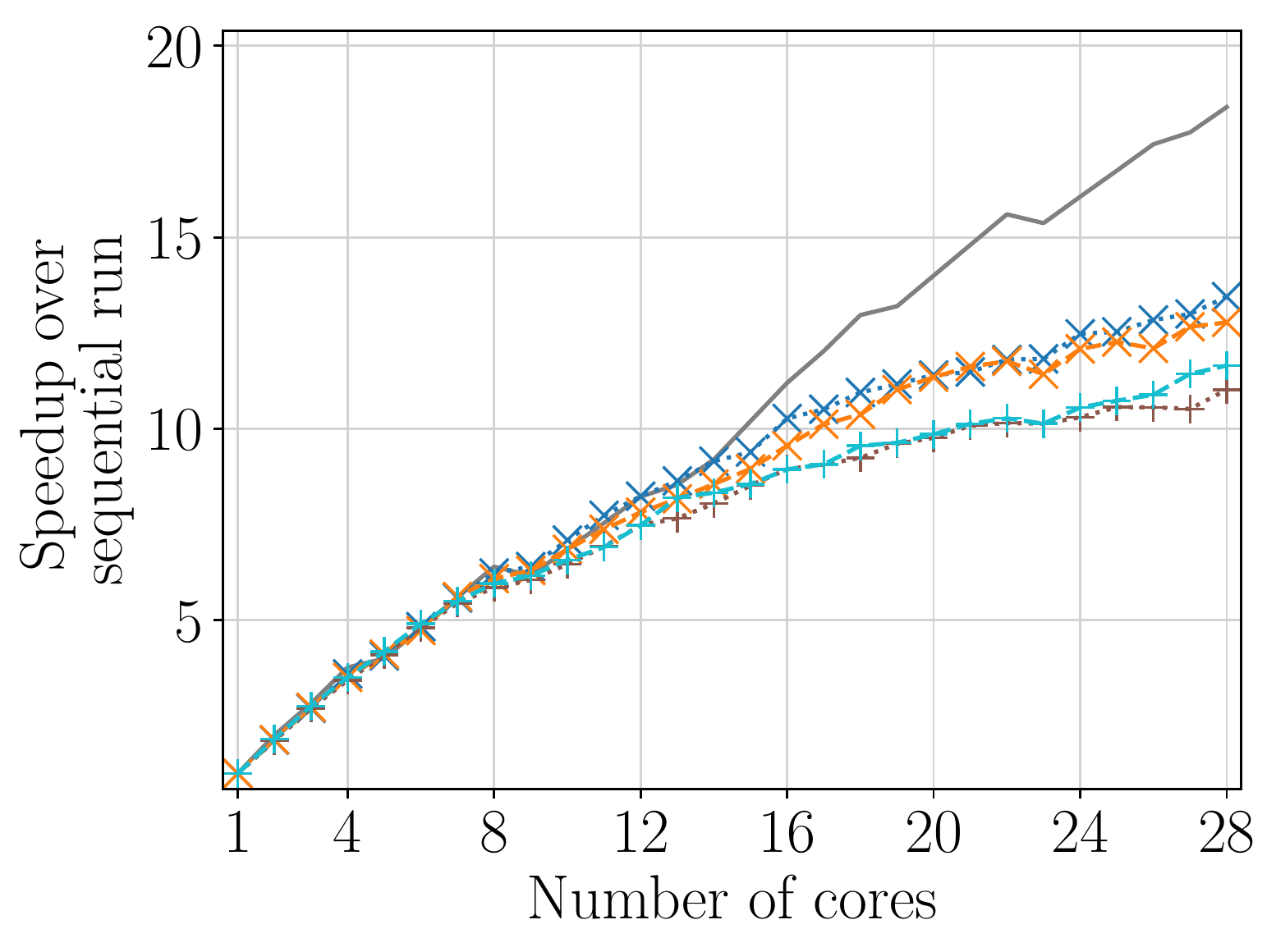}
  \end{minipage}
  \caption{Parallel scalability on $\boldsymbol{H}_1$ using $n = 10000$ (left column) and $n = 40000$ (right column).}
  \label{fig:parallel-scalability}
\end{figure}

\paragraph{Parallel scalability}
The third experiment analyzes the strong scalability of \textsc{HSRQ3IN}. Strong scaling concerns the speedup for a fixed problem size subject to an increasing number of processing units. The used test system is $\boldsymbol{H}_1$ where $n \in \{10000, 40000\}$. The eigenvalues are either all real or all complex. The eigenvalue selection ratio is chosen as $5\%$ or $25\%$. Then, for example, the experiment with $n=10000$ and $5\%$ selected eigenvalues computes 500 columns (500 real eigenvectors or 250 complex eigenvectors). The top row of Figure~\ref{fig:parallel-scalability} displays the performance results with respect to the machine capabilities. The plot assumes $3.5 n^2$ flops per column, which is a lower bound of the true flop count. The bottom row of Figure~\ref{fig:parallel-scalability} shows the parallel speedup. The real and the complex experiments attain a similar fraction of the theoretical peak performance and achieve a similar parallel speedup.

\paragraph{Overhead due to numerical scaling}
The fourth experiment evaluates the cost of overflow avoidance. For this purpose, a run of \textsc{DHSRQ3} on the bad system $\boldsymbol{H}_2$ and the good system $\boldsymbol{H}_3$ are compared. Recall that the bad system requires frequent numerical scaling, whereas the good system never requires numerical scaling. Note that the computation on the good system is essentially identical to the real computation of Figure~\ref{fig:analysis}.

\begin{figure}
  \begin{minipage}[b]{.375\textwidth}
    \includegraphics[width=\textwidth]{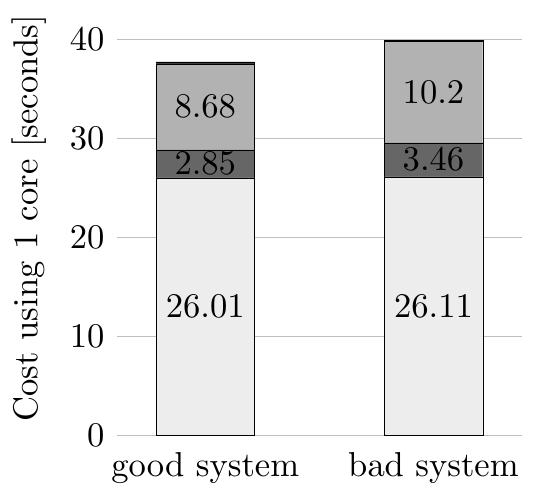}
  \end{minipage}
  \begin{minipage}[b]{.6\textwidth}
    \includegraphics[width=\textwidth]{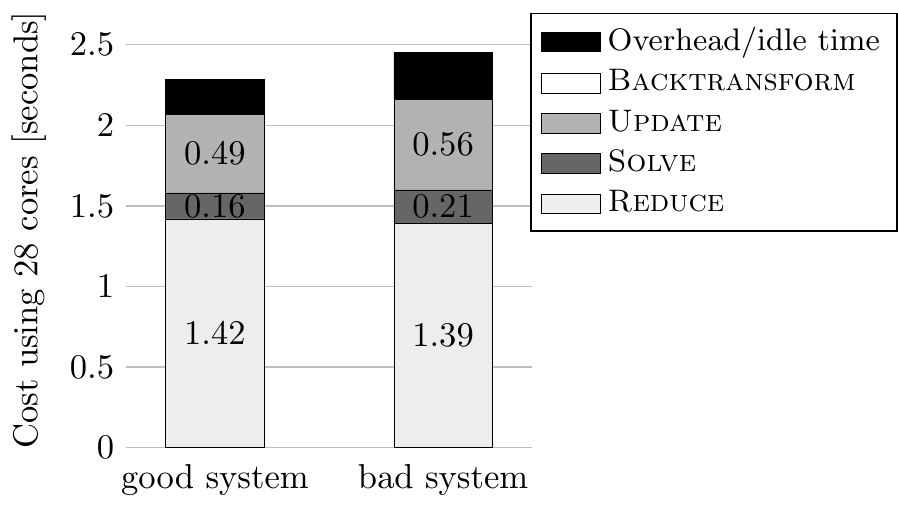}
  \end{minipage}
  \caption{Runtime decomposition of a sequential (left) and parallel (right) run computing 1500 real right-hand sides on $\boldsymbol{H}_2$ (bad system) and $\boldsymbol{H}_3$ (good system) for $n = 10000$.}
  \label{fig:numerical-scaling}
\end{figure}

Figure~\ref{fig:numerical-scaling} illustrates the measurements using $n = 10000$ and 1500 real right-hand sides. The runtime difference between the good and the bad system allows quantifying the overhead due to numerical scaling. The runs on the bad system are circa 6\% slower than the ones on the good system. Numerical scaling only affects the backward substitution phase, i.e., \textsc{Solve} and \textsc{Update} tasks. The slowdown of these two task types is approximately 20\% alike. The parallel speedup is comparable for both the good and the bad system. This suggests that robustness does not affect the parallel scalability.

\section{Conclusion and Outlook}

This paper revises the RQ approach for solving shifted Hessenberg systems $(\boldsymbol{H} - \lambda_{\ell}\boldsymbol{I}) \boldsymbol{x}_{\ell} = \alpha_{\ell} \boldsymbol{b}_{\ell}$ robustly for a large number of shifts. By rearranging the computation of the partial RQ factorization, matrix--matrix multiplications (level-3 BLAS) are introduced to the backward substitution phase. Since the solution of shifted Hessenberg systems is the most compute-intensive step in the computation of eigenvectors by inverse iteration, the revised RQ approach leads to a new inverse iteration solver. The numerical experiments show that the new inverse iteration solver outperforms existing inverse iteration solvers.

By improving the backward substitution phase, the reduction phase computing the orthogonal Q factor has become the largest contributor to the runtime. In view of Amdahl's law, a reasonable next step is investigating options to reduce the impact of the reduction phase. In particular ideas proposed for the original RQ approach such as improving the cache efficiency and aggregating a small number of orthogonal transformations for a joint application \cite[p. 8]{henry1994techreport}\cite[Sec. 5.2.1]{beattie2012note} can be scrutinized with respect to their usefulness in the new algorithm.

\section{Acknowledgements}
The author thanks Lars Karlsson for initiating and supporting this project. Furthermore, the author is grateful for the valuable comments by Martin Berggren, Lars Karlsson and Carl Christian Kjelgaard Mikkelsen. Computing resources have been provided by the Swedish National Infrastructure for Computing (SNIC) at High-Performance Computing Center North (HPC2N), Ume\r{a}, Sweden, under the grants SNIC 2019/3-311 and SNIC 2020/5-286.

\bibliographystyle{plain}
\bibliography{literature}

\end{document}